\documentclass[aps,prd,preprintnumbers,superscriptaddress,nofootinbib]{revtex4}%

\makeatletter
    
    \@addtoreset{equation}{section}
  \makeatother

\usepackage[dvipdfmx]{graphicx} 
\usepackage{feynmf}
\usepackage{color}
\usepackage{bm,latexsym,amsmath,amssymb,amsfonts,mathrsfs}
\usepackage[]{hyperref}

\usepackage{physics}	
\usepackage{xcolor}
\usepackage{tensor}
\usepackage{ulem}

\usepackage{physics}
\usepackage{aas_macros}

\newcommand*{\beqa}{\begin{eqnarray}}
\newcommand*{\eeqa}{\end{eqnarray}}

\newcommand*{\p}{\partial}

\newcommand{\A}{{\bf A}}

\newcommand{\bL}{{\boldsymbol{L}}}
\newcommand{\bC}{{\boldsymbol{C}}}
\newcommand{\ba}{{\boldsymbol{a}}}
\newcommand{\bb}{{\boldsymbol{b}}}

\newcommand{\x}{{\boldsymbol{x}}}
\newcommand{\bp}{{\boldsymbol{p}}}

\newcommand{\bphi}{{\boldsymbol{\phi}}}

\newcommand{\tilt}{\widetilde{t}}

\begin{document}

\title{Eisenhart Lift for Scalar Fields in the FLRW Universe}

\author{
Takeshi Chiba\footnote{Corresponding author; email: chibatak@gmail.com}
}
\affiliation{Department of Physics, College of Humanities and Sciences, Nihon University, \\
                Tokyo 156-8550, Japan}
\author{
Tsuyoshi Houri
}
\affiliation{National Institute of Technology, Maizuru College, \\
Kyoto 625-8511, Japan
}
\date{\today}

\begin{abstract}
The Eisenhart lift of Riemannian type describes the motion of a particle 
as a geodesic  in a  higher-dimensional Riemannian manifold with one 
additional coordinate.  
It has recently been generalized to a scalar field system by introducing one additional vector field. 
We apply this approach to a scalar field system in the Friedmann-Lemaitre-Robertson-Walker universe and classify the symmetries of the system. In particular, 
for a scalar field potential consisting of the square of a combination of  
exponential functions with specific index $e^{\pm\frac{\sqrt{6}}{4}\phi}$, 
we find nontrivial (conformal) Killing vector fields and Killing tensor fields. 
Moreover, for a potential written as an  exponentiation of a combination 
of exponential potentials with general index, 
we find  nontrivial conformal Killing vector fields.  
By introducing the coordinate along the conformal Killing vector field, 
we can solve the equations of motion completely.
We also classify the symmetries of multiple scalar field system. 
\end{abstract}


\maketitle

\section{Introduction}

Eisenhart showed that the classical motion of a particle under 
the influence of a potential $V(\x)$ is equivalent to a geodesic 
of a higher dimensional Riemannian manifold with one extra coordinate 
(which we call Eisenhart lift of the Riemannian type)\cite{eisenhart1928} (see \cite{Cariglia:2014ysa, Cariglia:2015bla} for a review). 
Eisenhart also introduced a different type of geometric lift in which 
the motion of a particle can be viewed as a null geodesic of a higher dimensional Lorentzian manifold with two extra coordinates (which we call Einsenhart lift of Lorentzain type)\cite{eisenhart1928, Cariglia:2014ysa,Cariglia:2015bla}. Finding the conserved quantities of 
the system is reduced to finding the conserved quantities for 
geodesic motion, which are determined by the Killing vectors (KVs) and Killing tensors (KTs) of the lifted metric.

The Eisenhart lift of the Riemannian type is extended 
to scalar field theories by introducing one vector field \cite{Finn:2018cfs}. 
We apply this approach to the system of a scalar field in the Friedmann-Lemaitre-Robertson-Walker (FLRW) universe 
and investigate the symmetry of the field space by studying 
the KV and conformal KV (CKV) fields. 
In our previous paper \cite{Chiba:2024iia}, we briefly outlined the method  and found that a CKV field exists for a particular combination of exponential potentials which included a single exponential potential and gave  
a particular solution of the equations of motion. In this paper, we provide a 
more detailed account of the symmetries of the system and solve 
the equations of motion explicitly to give complete solutions to the 
equations of motion.

In particular,  for $V(\phi)$ consisting of the square of a combination of  
exponential functions with specific index $e^{\pm\frac{\sqrt{6}}{4}\phi}$ (see (\ref{potential-I}) below) 
which was already studied in the literature \cite{deRitis:1990ba,Garay:1990re}, 
we find nontrivial KV fields.  
Hence, by introducing Killing coordinates along the KV fields, 
we can solve the equations of motion easily. 
In addition to  KV fields, we also find a nontrivial KT 
field for this potential, which is not known in the literature as far as we are aware. 
Moreover, for $V(\phi)$  written as an    
exponentiation of a combination of exponential potentials with 
general index  (see (\ref{potential-III}) below), 
we find a nontrivial CKV field.   This may  
reveal the ``hidden symmetry'' of power-law inflation \cite{Chiba:2024iia}. 
Furthermore, for a single exponential potential (see (\ref{potential-IV}) below) 
which includes the potential for 
power-law inflation \cite{Lucchin:1984yf},  we find that the field space is 
conformally flat and there exist ten CKVs. 
We provide the complete set of the CKVs.

The paper is organized as follows.
In Sec.~\ref{sec2},  after reviewing the Eisenhart lift of Riemannian type 
for a particle, we introduce Eisenhart lift of Riemannian type for 
scalar fields.  In Sec.~\ref{sec3}, the system of a homogeneous scalar field  
in FLRW universe is lifted, and the symmetries of 
the system are classified. 
In Sec.~\ref{appendixB}, the explicit solutions of the equation of motion 
of a scalar field and the scale factor in the FLRW universe 
for the potential admitting (C)KV fields are given. 
In Sec.~\ref{sec6}, we discuss some problems in lifting spacetime-dependent fields by 
introducing the vierbein \cite{Finn:2018cfs}. 
Sec.~\ref{sec7} is devoted to summary.   In Appendix \ref{appendixA}, we review the particle motion under the central force and show  
the existence of extra conserved quantity associated with a KT for Kepler potential or harmonic potential.  
In Appendix \ref{appendixC}, we describe the structure of the CKV equation 
and provide the integrability condition of 
the CKVs.  In Appendix~\ref{multi}, we apply the Eisenhart lift to a multiple-field system and 
study the condition for the existence of CKVs. 
In Appendix~\ref{sec4}, after reviewing the Eisenhart lift of Lorentzian type for a particle, 
we attempt to construct  the Eisenhart lift of Lorentzian type 
for scalar fields by introducing two vector fields  and point 
out the problems of the framework.

Our convention of the metric signature is $(-,+,+,+)$ and we use 
the units of $8\pi G=c=1$. 

\section{Eisenhart Lift of Riemannian type}
\label{sec2}

\subsection{Eisenhart Lift for a Particle}


Consider the equation of motion of a particle under 
the influence of a potential $V(x^1,\dots,x^n)$ in an $n$-dimensional space
\beqa
m \ddot x^{i}=-\p_{i}V\,,
\label{eom1}
\eeqa
where $i=1,\dots ,n$ and $\ddot x^i=d^2x^i/dt^2$ with $t$ 
being the time coordinate and $\p_iV=\p V/\p x^i$. 
Hereafter we set $m=1$ for simplicity. The equation of motion  
is derived from the action
\beqa
S=\int dt \left(\frac12 \delta_{ij}\dot x^i\dot x^j-V(x^1,\dots,x^n)\right)\,.
\label{action1}
\eeqa

Eisenhart showed that the same dynamics can be derived from the following action  by adding a new coordinate $y$ \cite{eisenhart1928}
\beqa
I_R=\int dt\left(\frac12 \delta_{ij}\dot x^i\dot x^j+\frac{1}{4V}\dot y^2\right)\,,
\label{action2}
\eeqa
where $V\neq 0$ is assumed. It is called an Eisenhart lift of the Riemannian type since the metric signature is positive (if $V>0$). 
The equation of $x^i$ and $y$ are
\beqa
\ddot x^i&=&-\frac{1}{4V^2}\dot y^2\p_iV\,,
\label{eom2-1}\\
p_y&=&\frac{\dot y}{2V}={\rm const}\,.
\label{eom2-2}
\eeqa
So, plugging Eq. (\ref{eom2-2}) into Eq. (\ref{eom2-1}) and 
setting $p_y=1$ reproduces Eq. (\ref{eom1}). Changing 
$p_y$ merely corresponds to the rescaling of the time coordinate. 

Introducing $n+1$ dimensional coordinate $x^A=(x^i,y)$, 
the action Eq.(\ref{action2}) may be viewed as that of a free particle in 
$n+1$-dimensional space:
\beqa
I_R=\int dt~ \frac12 G_{AB}\dot x^A\dot x^B\,,
\eeqa
where $G_{AB}$ is the field space metric with 
$G_{ij}=\delta_{ij}, G_{iy}=0$ and $G_{yy}=\frac{1}{2V}$. 
Viewed in this way, the motion of a particle can be regarded as 
the geodesics  on a higher dimensional Riemannian manifold with 
the metric $G_{AB}$  \cite{eisenhart1928}.  
This method is called Eisenhart lift.  The motion of a particle under 
central force  and the existence of a Killing tensor for a Newtonian 
potential or harmonic potential are reviewed in Appendix \ref{appendixA}. 

It is interesting to note that for $n=1$ the Ricci scalar $R(G)$ 
constructed from $G_{AB}$ is 
 given by the Schwarzian derivative of $W(x)=\int dx V(x)$ (``pre-potential'' of $V(\phi)$), $S(W)$, 
\beqa
R(G)=\frac{V''}{V}-\frac32 \left(\frac{V'}{V}\right)^2=\frac{W'''}{W'}-\frac32 \left(\frac{W''}{W'}\right)^2\equiv S(W)\,.
\eeqa
Thus, the Schwarzian derivative is curvature \cite{schwarzian}. 
The Ricci scalar $R(G)$ for general $n$ becomes 
$R(G)=\frac{D^2V}{V}-\frac32 \frac{(D V)^2}{V^2}$,  
where $D$ is the covariant derivative with respect to $G_{AB}$.

\subsection{Eisenhart Lift for Scalar Fields}
\label{sec2B}

The Eisenhart lift of the Riemannian type is extended to scalar field theories by \cite{Finn:2018cfs}. 
Consider the system of $n$ scalar fields in a four-dimensional spacetime 
(with the metric $g_{\mu\nu}$) 
\beqa
S=\int d^4x\sqrt{-g}\left(\frac12 R-\frac12 g^{\mu\nu}k_{IJ}(\bphi)\p_{\mu}\phi^I\p_{\nu}\phi^J
-V(\bphi)\right)\,,
\label{action:scalar}
\eeqa
where the first term is the Einstein-Hilbert term (in units of $8\pi G=1$) 
and $I=1,\dots,n$ is the field space index and $k_{IJ}(\bphi)$ is
 the scalar field space metric, while $\mu$ is the spacetime index. 
The Einstein equation and  the equation of motion of $\phi^I$ are given by
 \beqa
&& R_{\mu\nu}-\frac12 g_{\mu\nu}R=k_{IJ}\p_{\mu}\phi^I\p_{\nu}\phi^J-\frac12 g_{\mu\nu}\left(g^{\alpha\beta}k_{IJ}\p_{\alpha}\phi^I\p_{\beta}\phi^J
+2V \right)\,,
\label{einstein:scalar1}\\
&& \Box\phi^I+\Gamma^{I}_{JK}g^{\mu\nu}\p_{\mu}\phi^J\p_{\nu}\phi^K-k^{IJ}\p_JV=0\,,
\label{eom:scalar1}
 \eeqa
where $\Gamma^I_{JK}$ is the Christoffel symbol constructed from the 
field space metric $k_{IJ}$. 

Ref. \cite{Finn:2018cfs} observed that the same dynamics can be described by the 
following the Eisenhart lift through the introduction of the 
fictitious vector field $B^{\mu}$
\beqa
I_R=\int d^4x\sqrt{-g}\left(\frac12 R-\frac12 g^{\mu\nu}k_{IJ}(\bphi)\p_{\mu}\phi^I\p_{\nu}\phi^J
+\frac{1}{4V(\bphi)}\left(\nabla_{\mu}B^{\mu}\right)^2
\right)\,.
\label{action:scalar2}
\eeqa
The Einstein equation and  the equations of motion of $\phi^I$ and $B^{\mu}$ are
\beqa
&&R_{\mu\nu}-\frac12 g_{\mu\nu}R=k_{IJ}\p_{\mu}\phi^I\p_{\nu}\phi^J-
\frac12 g_{\mu\nu}\left(g^{\alpha\beta}k_{IJ}\p_{\alpha}\phi^I\p_{\beta}\phi^J\right)\nonumber\\
&&~~~~~~~~~~~~~~~~~~~~+2B_{(\mu}\p_{\nu)}\pi_B-g_{\mu\nu}B^{\alpha}\p_{\alpha}\pi_B-g_{\mu\nu}V\pi_B^2\,,
\label{einstein:scalar2}
\\
&&\Box\phi^I+\Gamma^{I}_{JK}g^{\mu\nu}\p_{\mu}\phi^J\p_{\nu}\phi^K-
\pi_B^2k^{IJ}\p_JV=0\,,
\label{eom:scalar2-1}
\\
&&\p_{\mu}\pi_B=0\,,
\label{eom:scalar2-2}
\eeqa
where 
\beqa
\pi_B=\frac{\nabla_{\mu}B^{\mu}}{2V}\,.
\label{piB}
\eeqa 
{}From Eq. (\ref{eom:scalar2-2}),  $\pi_B$ is a constant.  
Plugging this into Eq. (\ref{einstein:scalar2}) and Eq. (\ref{eom:scalar2-1}) and setting $\pi_B=1$ reproduces Eq. (\ref{einstein:scalar1}) and Eq. (\ref{eom:scalar1}).

\section{Scalar Field in FLRW Universe: Riemannian type}
\label{sec3}

We apply the Riemannian-type Eisenhart lift for scalar fields to 
the system of a single scalar field in the  FLRW universe \cite{Cariglia:2018mos}. 

\subsection{Field Space}
We consider  a flat FLRW universe 
\begin{align}
   g_{\mu\nu}dx^{\mu}dx^{\nu}=-N(t)^2dt^2+a(t)^2d\x^2 
\end{align}
where $N(t)$ is the lapse function and $a(t)$ is the scale factor. 
Assuming that a scalar field $\phi$ and a vector field $B^{\mu}$ are homogeneous, 
 the lifted system (\ref{action:scalar2}) reduces to a particle system 
governed by the Lagrangian\footnote{
It is interesting to note that in terms of new coordinates $x,y$ such that  \cite{Garay:1990re}
\beqa
x&=&\sqrt{8/3}a^{3/2}\cosh(\sqrt{3/8}\phi)\nonumber\,, \\
y&=&\sqrt{8/3}a^{3/2}\sinh(\sqrt{3/8}\phi)\nonumber\,,
\eeqa
 the kinetic terms are canonically normalized:
\beqa
\frac12\left(-6a\dot a+a^3\dot\phi^2\right)=
\frac12(-\dot x^2+\dot y^2)\nonumber\,.
\eeqa
Therefore, the system we are considering corresponds to the system 
on a two-dimensional Minkowski field space. 
Classification of the integrable system along the line of \cite{2001JPhA...34.4705K} would be very interesting.
}
 \beqa
 {\cal L}=-\frac{3a}{N}\dot a^2+\frac{a^3}{2N}\dot\phi^2+\frac{1}{4Na^3V}\dot\chi^2
 \equiv \frac12 G_{AB}\dot\varphi^A\dot\varphi^B\,,
 \label{L-FRW}
 \eeqa
where $\chi\equiv Na^3B^0$, $\varphi^A=(a,\phi,\chi)$, and the field space metric $G_{AB}$ is given by
\beqa
G_{AB}=
\begin{pmatrix}
-\frac{6a}{N} & &  \\
   & \frac{a^3}{N}&  \\
  &  &\frac{1}{2Na^3V} \\
\end{pmatrix}
\label{effective-metric}
\,.
\eeqa

Since we have $\frac{\partial {\cal L}}{\partial \chi}= 0$,
$p_\chi \equiv \frac{\partial {\cal L}}{\partial \dot{\chi}} 
= \frac{\dot{\chi}}{2Na^3 V}$ is constant,
which coincides with $\pi_B$ in Eq.~\eqref{piB}.
Here, we set $N=1$ for simplicity.
The Euler-Lagrange equations are given by
\begin{align}
 & \frac{\ddot a}{a}=-\frac{1}{3}\left(\dot\phi^2-Vp_\chi^2\right)\,,
 \label{Friedmann-Eq-1}\\
 & \left(\frac{\dot a}{a}\right)^2=\frac13\left(\frac12 \dot\phi^2+Vp_\chi^2\right) \,,
 \label{Friedmann-Eq-2}\\
 & \ddot\phi+3\frac{\dot a}{a}\dot\phi+V'p_\chi^2=0 \,,
 \label{Friedmann-Eq-3}
\end{align}
where $V'=dV/d\phi$. These equations are the same as those derived from the Lagrangian \eqref{action:scalar}
with $p_\chi=\pi_B=1$.

In terms of the conjugate momenta 
$p_A=(p_a,p_\phi,p_\chi)$, which are given by
$p_a=-6a\dot a/N$, $p_{\phi}=a^3\dot\phi/N$ 
and $p_{\chi}=\frac{\dot\chi}{2Na^3V}$,
the Hamiltonian is given by
\begin{align}
{\cal H}=\frac12G^{AB}p_Ap_B=N\left(-\frac{1}{12a}p_a^2+\frac{1}{2a^3}p_{\phi}^2+a^3V(\phi)p_{\chi}^2\right)\,,
\end{align}
from which the Hamiltonian constraint follows
\beqa
H= \frac{1}{2}\left(
-\frac{p_a^2}{6a}+\frac{p_\phi^2}{a^3}
+2a^3V(\phi)p_\chi^2 \right)=0\,.
\label{H-constraint}
\eeqa
Therefore, $p^A\equiv G^{AB}p_B$ is a null vector field. 
Henceforth, we set $N=1$.
The canonical equations of motion for $H$, 
$\dot\varphi^A=\frac{\p H}{\p p_A}$
and $\dot p_A=-\frac{\p H}{\p \varphi^A}$,
provide the equations for null geodesics on the field space,
\begin{align}
    \dot{a} =& -\frac{p_a}{6a} \,, \quad 
    \dot{p}_a = -\frac{p_a^2}{12a^2}
    +\frac{3p_\phi^2}{2a^4}
    -3a^2Vp_\chi^2 \,, 
    \label{eom-01}\\
    \dot{\phi} =& \frac{p_\phi}{a^3} \,, \quad
    \dot{p}_\phi = -a^3V'p_\chi^2 \,, 
    \label{eom-02}\\
    \dot{\chi} =& 2a^3Vp_\chi \,, \quad
    \dot{p}_\chi = 0 \,.
    \label{eom-03}
\end{align}
From the last equation, we find again that $p_\chi$ is a constant.
Along with the Hamiltonian constraint \eqref{H-constraint}
and $p_{\chi}=1$, these equations reproduce
Eqs.~\eqref{Friedmann-Eq-1}--\eqref{Friedmann-Eq-3}.

Let us compute the curvatures of the field space metric $G_{AB}$.
The Ricci tensor and scalar curvature are given by
\begin{align}
 R_{AB} =
\left(
\begin{array}{ccc}
-\frac{9}{2a^2} &-\frac{3V'}{2aV} &  \\
-\frac{3V'}{2aV} &\frac{\Phi(V)}{8}-\frac{3}{4} & \\
 & &\frac{\Phi(V)}{16a^6 V}
\end{array}
\right)
\end{align}
and
\begin{align}
R = \frac{\Phi(V)}{4a^3} \,,
\end{align}
where
\begin{align}
 \Phi(V) = 3+4\left[\frac{V''}{V}
 -\frac{3}{2}\left(\frac{V'}{V}\right)^2\right] \,. \label{def-Phi}
\end{align}
Again, the Ricci scalar is given by the Schwarzian derivative of the prepotential $W(\phi)=\int d\phi V(\phi)$, $S(W)$: $a^3R=\frac34 +S(W)$.\footnote{It is tempting to relate $R$ to the spectral index of the curvature perturbation generated during inflation \cite{Karananas:2023bog}. However, $R$ has nothing to do with inflation: 
even in a flat spacetime $R$ involves the Schwarzian derivative. Moreover,  
we do not assume slow-roll approximations in deriving $R$, 
but rather the expression of $R$ is exact.} 
In particular, when the potential is given by
\begin{align}
 V(\phi) = \left(c_1 e^{\frac{\sqrt{6}}{4}\phi}
 +c_2 e^{-\frac{\sqrt{6}}{4}\phi} \right)^{-2} \,,
\end{align}
where $c_1$ and $c_2$ are arbitrary constants,
the quantity $\Phi(V)$ vanishes and hence
the $\chi\chi$-component of the Ricci tensor and 
the scalar curvature $R$ vanish.

It is also interesting that, when the potential is given by
\begin{align}
 V(\phi) =  e^{\lambda \phi} \,,
\end{align}
where $\lambda$ is a constant, the Cotton tensor (see Appendix \ref{appendixC} for its definition) vanishes and
then the field space becomes conformally flat.

\subsection{(Hidden) Symmetries}
We explore the (C)KVs 
and (conformal) KT fields ((C)KTs) 
on the field space with the metric \eqref{effective-metric}.
An immediate consequence is that,
since the metric components do not depend on the coordinate $\chi$,
we always find the KV 
\begin{align}
  \xi_{(1)} = \partial/\partial \chi \,,
\end{align}
which is providing the constant of motion $p_\chi=\xi_{(1)}^Ap_A$.

Our interest here is whether other (C)KVs and/or (C)KTs exist
for some specific potential $V(\phi)$. 
This is resolved by solving (C)KV and (C)KT equations
and we find nontrivial (C)KVs and (C)KTs 
when the potential $V(\phi)$ is given by the following cases.

\subsubsection{General case}
For an arbitrary $V(\phi)$, the field space metric \eqref{effective-metric} admits
a proper CKV\footnote{The ``proper" means that it is not a KV field but 
a genuinely CKV field. }
\begin{equation}
 \xi_{(2)} = a\frac{\partial}{\partial a}
 + 3\chi \frac{\partial}{\partial \chi} \,,
\end{equation}
which provides a constant of motion along null geodesics,
\begin{equation}
 \ell = a p_a + 3\chi p_\chi \,.
 \label{com_ell}
\end{equation}
Note that $\xi_{(1)}$ and $\xi_{(2)}$ do not commute, 
$[\xi_{(1)},\xi_{(2)}]=3\xi_{(1)}$,
and three constants of motion are not in involution. Hence, the system is not integrable
in the Liouville sense.
However, using three constants of motion
$H=0$, $\ell$ and $p_\chi=1$, 
the equations of motion \eqref{eom-01}--\eqref{eom-03}
can be reduced to a system of first-order equations for
$a,\phi$ and $\chi$.
In fact, by using Eqs.~\eqref{H-constraint}, \eqref{com_ell}, and $p_\chi =1$,
the momenta $p_a$ and $p_\phi$ are given by
\begin{align}
 p_a =& \frac{\ell -3\chi}{a} \,, \\
 p_\phi =& \pm \sqrt{\frac{1}{6}(\ell-3\chi)^2
 -2a^6 V(\phi)} \,.
\end{align}
Substituting these expressions into the equations of motion,
we have
\begin{align}
 \dot{a} =& -\frac{\ell -3\chi}{6a^2} \,, \\
 \dot{\phi} =& \pm \frac{1}{a^3}
 \sqrt{\frac{1}{6}(\ell-3\chi )^2
 -2a^6 V(\phi)} \,, \\
 \dot{\chi} =& 2a^3 V(\phi) \,.
\end{align}

\subsubsection{Case I}
The first case (which we call Case I) is obtained 
by requiring the field space metric \eqref{effective-metric} 
to admit at least one KV other than $\xi_{(1)}$.
By solving the Killing vector equations,
the potential is given by
\begin{align}
 V(\phi) = V_0
 \left(\cos(\beta) e^{\frac{\sqrt{6}}{4}\phi} 
 + \sin(\beta) e^{-\frac{\sqrt{6}}{4}\phi}\right)^2 \,,
 \label{potential-I}
\end{align}
where $V_0$ is a nonzero constant and $\beta$ is a constant
ranging $0\leq \beta<\pi$.
We note that the form 
of the potential is {\it derived} by solving the Killing vector equations
rather than being assumed before solving the equations. 
 This potential is the most general one  
 admitting an additional Killing vector field.


In what follows, we assume that
$0<\beta<\pi/2$ or $\pi/2 <\beta<\pi$,
which are collectively expressed by $\sin(2\beta)\neq 0$.
Under this assumption, we find an additional KV
\begin{align}
 \xi_{(3)} =& -a^{-\frac{1}{2}}
 \left(\cos(\beta) e^{\frac{\sqrt{6}}{4}\phi} 
 - \sin(\beta) e^{-\frac{\sqrt{6}}{4}\phi}\right)
 \frac{\partial}{\partial a}
 + \sqrt{6}a^{-\frac{3}{2}}
 \left(\cos(\beta) e^{\frac{\sqrt{6}}{4}\phi} 
 + \sin(\beta) e^{-\frac{\sqrt{6}}{4}\phi}\right) 
 \frac{\partial}{\partial \phi} \,.
 \label{KV-caseI-1}
\end{align}
The potential (\ref{potential-I}) and the KV $\xi_{(3)}$ coincide with 
those studied by \cite{deRitis:1990ba} where the integrable 
cosmological models are studied by searching for the Noether symmetry 
for the Lagrangian. 
For $\sin(2\beta)=0$, since the field space is conformally flat, 
the field space admits more (C)KVs, as is described in the next subsection.

Since the commutation relations between $\xi_{(1)}$, $\xi_{(2)}$ and $\xi_{(3)}$ are given by
\begin{align}
 [\xi_{(1)}, \xi_{(3)}]=0 \,, \quad 
 [\xi_{(2)}, \xi_{(3)}]=-\frac{3}{2}\xi_{(3)} \,,
\end{align}
three constants of motion corresponding to $G_{AB}$,
$\xi_{(1)}$, and $\xi_{(3)}$ are in involution.

Besides the trivial KTs consisting of the metric and 
the symmetric tensor products of the KVs obtained above,
we find a nontrivial solution to the KT equation, which is given by 
\beqa
K^{AB}= \xi_{(2)}^{(A}\xi_{(3)}^{B)}
 +12a^{\frac{3}{2}}
 \left(\cos(\beta) e^{\frac{\sqrt{6}}{4}\phi} 
     - \sin(\beta) e^{-\frac{\sqrt{6}}{4}\phi}\right) 
 G^{AB}\,.
 \label{irreducible-KT}
\eeqa
However, since we are considering null geodesics, 
$H=\frac12G^{AB}p_Ap_B=0$, and hence 
this does not provide an additional constant of motion.

After all, together with the Hamiltonian constraint (\ref{H-constraint}), 
We end up with three functionally independent constants of motion,
and the system is completely integrable in the Liouville sense. 
The explicit solutions of the equations of motion are described in Sec.~\ref{appendixB-1}.

\subsubsection{Case II}

The second case (which we call Case II) is the case 
when the potential is given by
\begin{align}
 V(\phi) = V_0 e^{\mp \frac{\sqrt{6}}{2}\phi} \,,
 \label{potential-II}
\end{align}
where $V_0$ is a nonzero constant.
This potential is obtained from 
the Case-I potential \eqref{potential-I}
by setting $\beta=0$ or $\pi/2$.

In this case, we find two additional KVs other than $\xi_{(1)}$:
\begin{align}
 \xi_{(3)} =& \pm a^{-\frac{1}{2}}
 e^{\mp \frac{\sqrt{6}}{4}\phi} 
 \frac{\partial}{\partial a}
+ \sqrt{6} a^{-\frac{3}{2}}
 e^{\mp \frac{\sqrt{6}}{4}\phi} 
 \frac{\partial}{\partial \phi} \,, 
 \label{xi3-case-II} \\
 \xi_{(4)} =& \chi \xi_{(3)} 
 \mp\frac{8}{3}V_0\,a^{\frac{9}{2}} e^{\mp \frac{3\sqrt{6}}{4}\phi}
 \frac{\partial}{\partial \chi} \,.
 \label{xi4-case-II}
\end{align}
Since the potential \eqref{potential-II} is
the limiting case of \eqref{potential-I} with $\sin(2\beta)=0$,
the existence of $\xi_{(3)}$ 
is obvious; on the other hand, $\xi_{(4)}$ arises
as special one, which comes from the vanishing of
the scalar curvature, see \eqref{def-Phi}.
Moreover, the Cotton tensor is vanishing in the present case,
so the field space metric $G_{AB}$ is conformally flat.  
Since this case is included in the conformally flat case,
all CKVs in such a case can be found in Sec. \ref{sec325}.
The solutions of the equations of motion are 
given in Sec. \ref{appendixB-2}. 

It is also interesting that $\xi_{(3)}$ is a null KV.
The commutation relations between $\xi_{(1)}$, $\xi_{(3)}$
and $\xi_{(4)}$ are given by
\begin{align}
 [\xi_{(1)}, \xi_{(3)}]=0 \,, \quad
 [\xi_{(1)}, \xi_{(4)}]=\xi_{(3)} \,, \quad
 [\xi_{(3)}, \xi_{(4)}]=0 \,.
\end{align}

Since we have three KVs,
we obtain six rank-2 KTs as the symmetric tensor products of the KVs 
and the metric as the trivial KT. 
As before, the KT  \eqref{irreducible-KT} with $\sin(2\beta)=0$ does 
not provide a new constant of motion.
Hence, we have four functionally independent constants of motion in total, 
and the system is superintegrable.

\subsubsection{Case III}
The third case (Case III) is the case when the potential 
is given by
\begin{align}
V(\phi)= V_0 \left(
 \cos(\beta) e^{\alpha \phi}
 +\sin(\beta) e^{-\alpha \phi}\right)^{-2+\frac{\sqrt{6}}{\alpha}} \,,
 \label{potential-III}
\end{align}
where $V_0$ is a nonzero constant, 
$\beta$ is a constant ranging $0\leq \beta< 2\pi$,
and $\alpha$ is a parameter that determines
the power of the potential. 
If $\alpha=\frac{\sqrt{6}}{4}$, this potential reduces to
Case I potential \eqref{potential-I}.
Also, if $\alpha=\frac{\sqrt{6}}{2}$, this potential becomes constant.
Therefore, in the following we assume that $\alpha\neq\frac{\sqrt{6}}{4}$ 
and $\alpha\neq\frac{\sqrt{6}}{2}$. 
We again stress that the form 
of the potential is {\it derived} by solving the CKV equations 
rather than being assumed before solving the equations. \footnote{We note, however, that the potential may not be the most general one because in solving the conformal Killing vector equations we assumed the factorized form for the components of the conformal Killing vector (see Appendix 
in  \cite{Chiba:2024iia}) and the general solution was not obtained. }

In what follows, we assume that $\sin(2\beta)\neq 0$.
For this potential, we find a proper 
CKV $\xi_{(3)}$ in addition to 
the KV $\xi_{(1)}=\p/\p\chi$ \cite{Chiba:2024iia}
\begin{align}
 \xi_{(3)} = -a^{-\sqrt{6}\alpha +1}
 \left( \cos(\beta) e^{\alpha\phi} 
 - \sin(\beta) e^{-\alpha \phi} \right)
 \frac{\partial}{\partial a} 
 + \sqrt{6}a^{-\sqrt{6}\alpha}
 \left(\cos(\beta) e^{\alpha\phi} 
 + \sin(\beta) e^{-\alpha \phi}\right)
 \frac{\partial}{\partial \phi} \,,
\end{align}
which obeys the CKV equation
$\nabla_{(A}\xi_{B)} = f G_{AB}$ 
with the conformal factor
\begin{align}
 f = \sqrt{6}\left(\alpha-\frac{\sqrt{6}}{4}\right)
 a^{-\sqrt{6}\alpha}\left( \cos(\beta) e^{\alpha\phi} 
 - \sin(\beta) e^{-\alpha \phi} \right) \,.
\end{align}
Since $p_A$ is null,  $\xi_{(3)}^Ap_A$ is a constant of motion along null geodesics. 
Moreover, $\xi_{(1)}=\p/\p\chi$ and $\xi_{(3)}$ commute, 
$[\xi_{(1)}, \xi_{(3)}]=0$.  
Therefore, together with the Hamiltonian constraint Eq.~(\ref{H-constraint}), 
we have three mutually commuting, functionally independent constants of motion 
for the system with three degrees of freedom, and 
the system is still completely integrable in the Liouville sense \cite{arnold}.
The equations of motion are solved in Sec.~\ref{appendixB-3}. 

As already mentioned before, there exists a CKV 
$\xi_{(2)}=a\p/\p a+3\chi\p/\p\chi$ irrespective of the form of $V(\phi)$. 
Therefore, we have two  CKVs  
which is the maximum number of independent CKVs 
(see Appendix \ref{appendixC}).  

Note that  if $\alpha=\frac{\sqrt{6}}{4}$, the potential
Eq. (\ref{potential-III}) 
reduces to \eqref{potential-I} and then
the CKV
becomes a KV, namely, $f=0$.

We looked for a CKT with the same form as \eqref{irreducible-KT},
but we couldn't find such a CKT.
We have not yet investigated the integrability conditions
of the CKT equation, so we do not know if a CKT exists in this case. 

\subsubsection{Case IV}
\label{sec325}
The fourth case (Case IV) is the case when the potential is given by
\begin{align}
 V(\phi) = V_0 e^{\mp (\sqrt{6}-2\alpha) \phi} \,,
 \label{potential-IV}
\end{align}
which is obtained from the Case III potential \eqref{potential-III}
with $\sin(2\beta)= 0$.  This form of potential is widely used in cosmology in 
the context of inflation \cite{Lucchin:1984yf} and dark energy \cite{Copeland:1997et}.
In this case, the Cotton tensor 
vanishes and the field space becomes conformally flat which was noticed in 
\cite{Chiba:2024iia}.
Then, we obtain the maximal number, namely ten, of CKVs (see Appendix \ref{appendixC} and \cite{Batista:2017wcm} for details) which was not provided in \cite{Chiba:2024iia}.
Solving the CKV equations, we obtain
\begin{align}
  \xi_{(1)} =& \frac{\partial}{\partial \chi} \,, \\
  \xi_{(2)} =& a\frac{\partial}{\partial a}
              +3\chi \frac{\partial}{\partial \chi} \,, \\
  \xi_{(3)} =& -a^{-\sqrt{6}\alpha+1}e^{\alpha\phi} \frac{\partial}{\partial a}
              + \sqrt{6} a^{-\sqrt{6}\alpha}e^{\alpha\phi} \frac{\partial}{\partial \phi} \,, \\
  \xi_{(4)} =& \chi\,\xi_{(3)}
              + 2\sqrt{6}V_0 a^{-\sqrt{6}\alpha+6} 
              e^{-(\alpha-\sqrt{6})\phi} \frac{\partial}{\partial \chi} \,, \\
  \xi_{(5)} =& -(\alpha-\sqrt{6})^2\chi^2 \,\xi_{(3)}
             +2 V_0 a^{-\sqrt{6}\alpha+6}e^{(\sqrt{6}-\alpha)\phi}
              \left( a\frac{\partial}{\partial a}+\sqrt{6}\frac{\partial}{\partial\phi}
              -2\sqrt{6} (\alpha-\sqrt{6})\chi\frac{\partial}{\partial\chi} \right) \,, \\
  \xi_{(6)} =& a^{\sqrt{6}\alpha-5}e^{(\alpha-\sqrt{6})\phi} \frac{\partial}{\partial a} 
              +\sqrt{6} a^{\sqrt{6}\alpha-6}e^{(\alpha-\sqrt{6})\phi} \frac{\partial}{\partial \phi} \,, \\
  \xi_{(7)} =& \alpha \chi \,\xi_{(6)}
              +2\sqrt{6} V_0 a^{\sqrt{6}\alpha}e^{-\alpha\phi}
              \frac{\partial}{\partial \chi} \,, \\
  \xi_{(8)} =& \alpha^2 \chi^2 \,\xi_{(6)}
              + 2V_0 a^{\sqrt{6}\alpha} e^{-\alpha\phi}
              \left( a\frac{\partial}{\partial a}
              - \sqrt{6}\frac{\partial}{\partial \phi} 
              + 2\sqrt{6}\alpha \chi \frac{\partial}{\partial \chi}
              \right) \,, \\
  \xi_{(9)} =& a\chi \frac{\partial}{\partial a} 
               + (2\alpha-\sqrt{6})\chi \frac{\partial}{\partial \phi}
              + \left(2V_0 a^6 e^{-(2\alpha-\sqrt{6})\phi}- \alpha(\alpha-\sqrt{6}) \chi^2 \right)
              \frac{\partial}{\partial \chi}  \,, \\
  \xi_{(10)} =& (2\alpha-\sqrt{6}) a \frac{\partial}{\partial a}
              + 6 \frac{\partial}{\partial \phi} \,,
\end{align}
where $\xi_{(1)}$ is a KV 
and $\xi_{(2)}$--$\xi_{(10)}$ are proper CKVs.
If $\alpha=\sqrt{6}/4$, the potential becomes \eqref{potential-II} in Case II.
Then $\xi_{(3)}$ and $\xi_{(4)}$ become KVs,
although $\xi_{(5)}$--$\xi_{(10)}$ do not.
If $\alpha=\sqrt{6}/2$, the potential becomes a constant
and then $\xi_{(10)}$ becomes a KV $\partial/\partial \phi$. 
The equations of motion are solved in Sec.~\ref{appendixB-4}.

\section{Exact Solutions}
\label{appendixB}

In this section, we solve the equations of motion
\eqref{Friedmann-Eq-1}--\eqref{Friedmann-Eq-3} 
with the potentials in Case I--IV
by utilizing the Riemannian-type Eisenhart lift
to consider the equations of motion as null geodesic equations on a field space
and by utilizing symmetries of a field space such as (C)KVs.
In the previous section, we have obtained a (C)KV $\xi_{(3)}$
which commutes with a KV $\xi_{(1)}=\partial/\partial \chi$ as $[\xi_{(1)}, \xi_{(3)}] = 0$.
Using it in this section, we choose a coordinate $\bar{\phi}$
whose coordinate basis $\partial/\partial \bar{\phi}$
coincides with the KV $\xi_{(3)}$,
i.e., $\xi_{(3)} = \partial/\partial \bar{\phi}$.
Then the components of the field space do not depend on the coordinates $\bar{\phi}$
and $\chi$, which makes the equations of motion simple and easy to solve.\footnote{
In \cite{Chiba:2024iia} we performed the canonical transformation to 
solve the equations of motion.  However, we find it more advantageous to 
adopt the Killing coordinate $\bar{\phi}$ to solve the equations of motion completely.}

\subsection{Solutions: Case I}
\label{appendixB-1}

In Case I, we consider the potential \eqref{potential-I}.
Then, if we choose a coordinate transformation\footnote{The transformation is derived by demanding $\xi_{(3)}=\left(\xi_{(3)}^a\frac{\p \bar{a}}{\p a}+\xi_{(3)}^{\phi}\frac{\p \bar{a}}{\p \phi}\right)\frac{\p}{\p \bar{a}}+\left(\xi_{(3)}^a\frac{\p \bar{\phi}}{\p a}+\xi_{(3)}^{\phi}\frac{\p \bar{\phi}}{\p \phi}\right)\frac{\p}{\p \bar{\phi}}=\frac{\p}{\p \bar{\phi}}$. This determines $\bar{\phi}(a,\phi)$, and $\bar{a}(a,\phi)$ is fixed (up to a constant factor) so that the dependence on $a$ has the same as $\bar{\phi}$.  }
\begin{align}
 \bar{a} =& a^{\frac{3}{2}}\left(
 \cos(\beta) e^{\frac{\sqrt{6}}{4}\phi}
 + \sin(\beta) e^{-\frac{\sqrt{6}}{4}\phi} \right) \,, \\
 \bar{\phi} =& \frac{1}{3}a^{\frac{3}{2}} 
 \left(
 \sin(\beta) e^{\frac{\sqrt{6}}{4}\phi}
 - \cos(\beta) e^{-\frac{\sqrt{6}}{4}\phi} \right) \,,
\end{align}
or, equivalently,
\begin{align}
 a=& \Big\{ \left(\cos(\beta)\bar{a}+3\sin(\beta)\bar{\phi}\right)
 \left(\sin(\beta)\bar{a}-3\cos(\beta)\bar{\phi}\right) \Big\}^{\frac{1}{3}}
  \,, \label{coord-transf-case-Ia} \\
 \quad
 \phi =& \frac{\sqrt{6}}{3} \ln
 \Bigg[ \frac{\cos(\beta)\bar{a}+3\sin(\beta)\bar{\phi}}
 {\sin(\beta)\bar{a}-3\cos(\beta)\bar{\phi}} \Bigg] \,,
 \label{coord-transf-case-Ib}
\end{align}
the field space metric \eqref{effective-metric} becomes
\begin{align}
G_{AB} = 
\left(
\begin{array}{ccc}
-\frac{4}{3}\sin(2\beta) &4\cos(2\beta) &  \\
4\cos(2\beta) &12\sin(2\beta) &  \\
 & &\displaystyle{\frac{1}{2\bar{a}^2V_0}}
\end{array}
\right) \,.
\label{effective-metric-trsf1}
\end{align}

The Hamiltonian is given by
\begin{align}
 H = -\frac{3\sin(2\beta)}{8} p_{\bar a}^2
 +\frac{\cos(2\beta)}{4}p_{\bar a}p_{\bar \phi}
 +\frac{\sin(2\beta)}{24}p_{\bar \phi}^2 
 + \bar{a}^2V_0p_\chi^2 \,,
 \label{Hamiltonian:B}
\end{align}
and the equations of motion are given by
\begin{align}
 \dot{\bar{a}} =& -\frac{3\sin(2\beta)}{4} p_{\bar a}
 +\frac{\cos(2\beta)}{4}p_{\bar \phi} \,,
 \qquad \dot{p}_{\bar a} =-2 \bar{a} V_0 p_\chi^2 \,,
 \label{eom-11}\\
 \dot{\bar{\phi}} =& \frac{\sin(2\beta)}{12}p_{\bar \phi}
 +\frac{\cos(2\beta)}{4}p_{\bar a} \,,
 \qquad \dot{p}_{\bar \phi} = 0 \,, 
 \label{eom-12}\\
 \dot{\chi} =& 2\bar{a}^2 V_0 p_\chi \,,
 \qquad \dot{p}_\chi = 0 \,.
 \label{eom-13}
\end{align}
In the current case, since we have $\sin(2\beta) \neq 0$,
the Hamiltonian constraint $H=0$ is solved by
\begin{align}
 p_{\bar a} = 
 \frac{\cos(2\beta)p_{\bar \phi}
 \pm \sqrt{
 p_{\bar \phi}^2
 +\epsilon \mu^2 \bar{a}^2}
 }{3 \sin(2\beta)} \,,
 \label{pa-case-I}
\end{align}
where $\mu^2=24 \epsilon \sin(2\beta)V_0 p_\chi^2$.
Here, we set $\epsilon=1$ for $\sin(2\beta)V_0>0$ or 
$\epsilon=-1$ for $\sin(2\beta)V_0<0$,
namely, $\mu$ is always chosen to be a real number.
We also find that $p_{\bar\phi}$ and $p_\chi$ are constants.
Substituting \eqref{pa-case-I} into the equations for $\bar{a}$ and $\bar{\phi}$,
we obtain
\begin{align}
\dot{\bar{a}} =& 
 \mp \frac{1}{4}\sqrt{
  p_{\bar \phi}^2
 +\epsilon \mu^2 \bar{a}^2 } \,, \\
 \dot{\bar{\phi}} =& 
 \frac{p_{\bar \phi}}{12\sin(2\beta)}
 \Bigg( 1 \pm \cos(2\beta) \sqrt{ 1
 +\epsilon (\mu^2/p_{\bar\phi}^2) \bar{a}^2} \Bigg) \,.
\end{align}
These equations can be integrated by using the hyperbolic sine function
for $\epsilon=1$ or the sine function for $\epsilon=-1$.

For $\epsilon=1$, the solutions to Eqs.~\eqref{eom-11}--\eqref{eom-13}
are given by
\begin{align}
 \bar{a} =&
 \mp \frac{p_{\bar \phi}}
 {\mu}\sinh
 \left[\frac{\mu}{4} (t-t_0) \right] \,, \\
 \bar{\phi} =& 
 \frac{p_{\bar \phi}}{12\sin(2\beta)} 
\left((t-t_0) \pm \frac{4}{\mu}\cos(2\beta) \sinh
\left[\frac{\mu}{4} (t-t_0) \right] \right) + \bar{\phi}_0 \,, \\
\chi =& -\frac{V_0p_{\bar{\phi}}^2p_\chi}{\mu^2}
\left((t-t_0)-\frac{2}{\mu}\sinh
\left[\frac{\mu}{2}(t-t_0)\right]
\right) + \chi_0 \,,
\end{align}
where $p_{\bar \phi}$, $p_\chi$,
$t_0$, $\bar{\phi}_0$, and $\chi_0$ are constants
and hence the solutions include five constants.
Moreover, by using the coordinate transformation \eqref{coord-transf-case-Ib}
with the redefinition of the constant $\bar{\phi}_0$
by $\bar{t}_0 = 12 \sin(2\beta) \bar{\phi}_0/p_{\bar{\phi}}$,
 the solutions to Eqs.~\eqref{Friedmann-Eq-1}--\eqref{Friedmann-Eq-3} are given by
\begin{align}
 a =& \left\{\frac{p_{\bar \phi}^2}{32\sin(2\beta)}
 \left(\frac{16}{\mu^2}\sinh^2\left[\frac{\mu}{4} (t-t_0) \right]
 -(t-t_0+\bar{t}_0)^2\right)\right\}^{\frac{1}{3}} \,, 
 \label{eq:IV16}\\
 \phi =& \frac{\sqrt{6}}{3}\ln\Bigg[-\tan(\beta)
 \frac{(t-t_0+\bar{t}_0)\mp (4/\mu)\sinh[(\mu/4)(t-t_0)]}
 {(t-t_0+\bar{t}_0)\pm (4/\mu)\sinh[(\mu/4)(t-t_0)]}\Bigg] \,.
\end{align}

For $\epsilon=-1$, the solutions to Eqs.~\eqref{eom-11}--\eqref{eom-13}
are given by
\begin{align}
 \bar{a} =&
 \mp \frac{p_{\bar \phi}}
 {\mu}\sin
 \left[\frac{\mu}{4} (t-t_0) \right] \,, \\
 \bar{\phi} =& 
 \frac{p_{\bar \phi}}{12\sin(2\beta)} 
\left((t-t_0) \pm \frac{4}{\mu}\cos(2\beta) \sin
\left[\frac{\mu}{4} (t-t_0) \right]
\right) +\bar{\phi}_0\,,\\
\chi =&
\frac{V_0p_{\bar{\phi}}^2p_\chi}{\mu^2}
\left((t-t_0)-\frac{2}{\mu}\sin
\left[\frac{\mu}{2}(t-t_0)\right]
\right) + \chi_0 \,,
\end{align}
where $p_{\bar \phi}$, $p_\chi$,
$t_0$, $\bar{\phi}_0$, and $\chi_0$ are constants
and hence the solutions include five constants.
Moreover, by using the coordinate transformation \eqref{coord-transf-case-Ib}
with the redefinition of the constant $\bar{\phi}_0$
by $\bar{t}_0 = 12 \sin(2\beta) \bar{\phi}_0/p_{\bar{\phi}}$,
 the solutions to Eqs.~\eqref{Friedmann-Eq-1}--\eqref{Friedmann-Eq-3} are given by
\begin{align}
 a =& \left\{ \frac{p_{\bar \phi}^2}{32\sin(2\beta)}
 \left(\frac{16}{\mu^2}\sin^2\left[\frac{\mu}{4} (t-t_0) \right]
 -(t-t_0+\bar{t}_0)^2\right) \right\}^{\frac{1}{3}} \,, 
 \label{eq:IV21}\\
 \phi =& \frac{\sqrt{6}}{3}\ln\left[-\tan(\beta)
 \frac{(t-t_0+\bar{t}_0)\mp (4/\mu)\sin[(\mu/4)(t-t_0)]}
 {(t-t_0+\bar{t}_0)\pm (4/\mu)\sin[(\mu/4)(t-t_0)]}\right] \,.
\end{align}
Eq. (\ref{eq:IV16}) and Eq. (\ref{eq:IV21}) show that the constant of motion associated with $\xi_{(3)}$, $p_{\bar{\phi}}$, corresponds to 
the normalization of the scale factor.

\subsection{Solutions: Case II}
\label{appendixB-2}

In Case II, we consider the potential \eqref{potential-II},
which is obtained by $\beta=0$ or $\pi/2$ from the Case-I potential.
We consider the coordinate transformation
\begin{align}
 \bar{a} =& a^{\frac{3}{2}}e^{\mp\frac{\sqrt{6}}{4}\phi} \,, \\
 \bar{\phi} =& \pm\frac{1}{3} a^{\frac{3}{2}} e^{\pm\frac{\sqrt{6}}{4}\phi} \,,
\end{align}
or, equivalently,
\begin{align}
 a =& \Big( \pm3\bar{a}\bar{\phi}\Big)^{\frac{1}{3}} \,, \\
 \phi =& \pm\frac{\sqrt{6}}{3}
 \ln \left[ \pm\frac{3\bar{\phi}}{\bar{a}}\right] \,.
\end{align}
Then, the CKVs \eqref{xi3-case-II} and \eqref{xi4-case-II} are given by $\xi_{(3)}=\partial/\partial \bar{\phi}$ and
\begin{align}
\xi_{(4)} = \chi \frac{\partial}{\partial \bar{\phi}}\pm
\frac{8}{3}V_0 \bar{a}^3 \frac{\partial}{\partial \chi} \,,
\end{align}
and the field space metric \eqref{effective-metric} becomes
\begin{align}
G_{AB} =
\left(
\begin{array}{ccc}
 &-4 &  \\
-4 & &  \\
 & &\displaystyle{\frac{1}{2\bar{a}^2V_0}} 
\end{array}
\right) \,.
\label{effective-metric-trsf21}
\end{align}

The Hamiltonian is given by
\begin{align}
 H = \mp\frac{1}{4}p_{\bar a} p_{\bar \phi}
 + V_0 \bar{a}^2p_\chi^2 \,,
 \label{hamiltonian-case2}
\end{align}
and the equations of motion are given by
\begin{align}
 \dot{\bar{a}} =& \mp\frac{1}{4}p_{\bar \phi} \,,
 \qquad \dot{p}_{\bar a} =-2 \bar{a} V_0p_\chi^2 \,,
 \label{eom-21}\\
 \dot{\bar{\phi}} =& \mp\frac{1}{4}p_{\bar a} \,,
 \qquad \dot{p}_{\bar \phi} = 0 \,, 
 \label{eom-22}\\
 \dot{\chi} =& 2 \bar{a}^2 V_0p_\chi \,,
 \qquad \dot{p}_\chi = 0 \,.
 \label{eom-23}
\end{align}
The equations can be easily solved by
\begin{align}
 \bar{a} =& \mp\frac{1}{4}p_{\bar \phi}(t-t_0) \,, \\
 \quad {\bar \phi} =& -\frac{1}{48}V_0p_{\bar \phi}p_\chi^2 (t-t_0)^3 + \bar{\phi}_0 \,, \\
 \quad \chi =& \frac{1}{24} V_0 p_{\bar \phi}^2 p_\chi (t-t_0)^3 + \chi_0 \,,
\end{align}
where $p_{\bar\phi}$, $p_\chi$, $t_0$, 
$\bar{\phi}_0$, and $\chi_0$
are constants.
The solutions to Eqs.~\eqref{Friedmann-Eq-1}--\eqref{Friedmann-Eq-3} are given by
\begin{align}
 a =& \left(\frac{1}{64}V_0p_{\bar \phi}^2
 p_\chi^2(t-t_0)^4
- \frac{3}{4}p_{\bar{\phi}}\bar{\phi}_0(t-t_0)\right)^{1/3}\,, \\
 \phi =& \pm\frac{\sqrt{6}}{3} \ln
 \left[ \frac{1}{4}V_0p_\chi^2 (t-t_0)^2-\frac{12\bar{\phi}_0}{p_{\bar{\phi}}(t-t_0)}\right] \,,
 \label{pl-sol-1}
\end{align}
reproducing the well-known solutions for 
for power-law inflation \cite{Lucchin:1984yf} for $\bar{\phi}_0=0$. 
$\bar{\phi}_0$ parameterizes the deviation from the power-law solution.

\subsection{Solutions: Case III}
\label{appendixB-3}

In Case III, we consider the potential \eqref{potential-III}.
We introduce the coordinate transformation
\begin{align}
 \bar{a} =& a^{\sqrt{6}\alpha}
 \Big(\cos(\beta) e^{\alpha\phi} + \sin(\beta) e^{-\alpha\phi}\Big) \,, \\
 \bar{\phi} =& \frac{a^{\sqrt{6}\alpha}}{2\sqrt{6}\alpha}
 \Big(\sin(\beta) e^{\alpha\phi} - \cos(\beta) e^{-\alpha\phi}\Big) 
 \,,
\end{align}
or, equivalently,
\begin{align}
 a =&
 \Big\{ \Big(\cos(\beta) \bar{a}+2\sqrt{6}\alpha \sin(\beta) \bar{\phi}\Big)
    \Big(\sin(\beta) \bar{a}-2\sqrt{6}\alpha \cos(\beta) \bar{\phi}\Big) 
    \Big\}^{\frac{1}{2\sqrt{6}\alpha}} \label{coordinate:a}\,, \\
 \phi =& \frac{1}{2\alpha}
 \ln\left[\frac{\cos(\beta) \bar{a}+2\sqrt{6}\alpha \sin(\beta) \bar{\phi}}
 {\sin(\beta) \bar{a}-2\sqrt{6}\alpha \cos(\beta) \bar{\phi}}\right] \label{coordinate:phi}\,.
\end{align}
Under this coordinate transformation, 
we have $\xi_{(3)}=\partial/\partial {\bar \phi}$
and the field space metric is
\begin{align}
G_{AB} = \Omega^{-1+\frac{\sqrt{6}}{4\alpha}} 
\left(
\begin{array}{ccc}
-\frac{1}{2\alpha^2}\sin(2\beta) &\frac{\sqrt{6}}{\alpha}\cos(2\beta) &  \\
\frac{\sqrt{6}}{\alpha}\cos(2\beta) &12\sin(2\beta) &  \\
 & &\displaystyle{\frac{\bar{a}^{2-\frac{\sqrt{6}}{\alpha}}}{2V_0}}
\end{array}
\right) \,.
\label{effective-metric-trsf3}
\end{align}

The Hamiltonian is given by
\begin{align}
 H = \Omega^{1-\frac{\sqrt{6}}{4\alpha}} H_0 \,,
 \label{H:H0}
\end{align}
where
\begin{align}
 H_0 =& -\alpha^2 \sin(2\beta) p_{\bar a}^2
 +\frac{\alpha\cos(2\beta)}{\sqrt{6}} p_{\bar a} p_{\bar \phi}
 +\frac{\sin(2\beta)}{24}p_{\bar \phi}^2 
 + \bar{a}^{-2+\frac{\sqrt{6}}{\alpha}} V_0 p_{\chi}^2 \,, \\
 \Omega =& \Big(\cos(\beta) \bar{a}+2\sqrt{6}\alpha \sin(\beta) \bar{\phi}\Big)
 \Big(\sin(\beta) \bar{a}-2\sqrt{6}\alpha \cos(\beta) \bar{\phi}\Big) \,.
\end{align}
Eqs.~(\ref{effective-metric-trsf3}) and (\ref{H:H0}) imply that 
$H_0$ is determined by the metric $G_{(0)AB}=\Omega^{1-\frac{\sqrt{6}}{4\alpha}}G_{AB}$. It is well known that  null geodesics are conformally 
invariant, namely, the null geodesics with respect to $G_{(0)AB}$ coincide with those 
with respect to $G_{AB}$ with the reparameterized affine parameter 
(corresponding to $t$ in our case) \cite{wald}. Specifically, once we obtain 
the solutions to the equations of motion with $H_0$, 
we immediately obtain the solutions with $H$
by changing the time coordinate from $\tilt$ to $t$ 
by $dt=\Omega^{-1+\frac{\sqrt{6}}{4\alpha}} d\tilt$. It is to be noted that   
$p_A$ is invariant under the conformal transformation\cite{wald,Chiba:2020mte} so that $H=\frac12 G^{AB}p_Ap_B=\frac12 \Omega^{1-\frac{\sqrt{6}}{4\alpha}}G_{(0)}^{AB}p_Ap_B=\Omega^{1-\frac{\sqrt{6}}{4\alpha}}H_0$. 

Thus, the equations to be solved here are the equations of motion with $H_0$:
\begin{align}
 & \dot{\bar a} =
 -2\alpha^2 \sin(2\beta) p_{\bar a}
 +\frac{\alpha \cos(2\beta)}{\sqrt{6}}p_{\bar \phi} \,, 
 \qquad \dot{p}_{\bar{a}} =
  - \left( -2+\frac{\sqrt{6}}{\alpha} \right)
 \bar{a}^{-3+\frac{\sqrt{6}}{\alpha}} V_0 p_\chi^2 \,, 
 \label{eom-case3-1}\\
 &\dot{\bar \phi} =
 \frac{\alpha \cos(2\beta)}{\sqrt{6}}p_{\bar a}
 +\frac{\sin(2\beta)}{12}p_{\bar \phi} \,,
 \qquad \dot{p}_{\bar \phi} =0 \,,
 \label{eom-case3-2}\\
 &\dot{\chi} = 
 2 \bar{a}^{-2+\frac{\sqrt{6}}{\alpha}} V_0 p_{\chi} \,,
 \qquad \dot{p}_\chi = 0 \,,
 \label{eom-case3-6}
\end{align}
where the dot denotes the differentiation with respect to
the affine parameter $\tilt$ of $H_0$.





From $H_0=0$, we have
\begin{align}
 p_{\bar a} = 
 \frac{\cos(2\beta)p_{\bar \phi}
 \pm \sqrt{
 p_{\bar \phi}^2
 +\epsilon \mu^2 \bar{a}^{-2+\frac{\sqrt{6}}{\alpha}}}
 }{2\sqrt{6}\alpha \sin(2\beta)} \,,
\end{align}
where we have set $\mu^2=24\epsilon \sin(2\beta) V_0$, again.
Hence, the equations for $\bar{a}$ and $\bar{\phi}$ become
\begin{align}
\dot{\bar{a}} =& 
 \mp \frac{\alpha p_{\bar\phi}}{\sqrt{6}}\sqrt{
  p_{\bar \phi}^2
 +\epsilon (\mu^2/p_{\bar\phi}^2) \bar{a}^{-2+\frac{\sqrt{6}}{\alpha}} } \,, \\
 \dot{\bar{\phi}} =&
 \frac{p_{\bar \phi}}{12\sin(2\beta)} 
 \left(1\pm \cos(2\beta)\sqrt{ 1
 +\epsilon (\mu^2/p_{\bar\phi}^2) \bar{a}^{-2+\frac{\sqrt{6}}{\alpha}}} \right)  \,.
\end{align}
By carrying out the integration, we obtain
\begin{align}
    \bar{a} =& p_{\bar\phi} F^{-1}(\tilt-\tilt_0) \,, \\
    \bar{\phi} =& \frac{p_{\bar \phi}}{12 \sin(2\beta)}\left(
    (\tilt-\tilt_0) - \frac{\sqrt{6}}{\alpha} \cos(2\beta) F^{-1}(\tilt-\tilt_0) \right) + \bar{\phi}_0 \,,
\end{align}
where $F^{-1}(z)$ is the inverse function of the function $F(z)$ given by
\begin{align}
 F(z) =& \int \frac{dz}{\sqrt{1+(\epsilon\mu^2/
 p_{\bar\phi}^2)z^{-2+\frac{\sqrt{6}}{\alpha}}}} \nonumber \\
 =& \pm \frac{\sqrt{6}\alpha^2
 z^{3-\frac{\sqrt{6}}{\alpha}}}{3(\sqrt{6}-4\alpha) \epsilon \mu^2}
 \sqrt{1+(\epsilon\mu^2/p_{\bar \phi}^2) z^{-2+\frac{\sqrt{6}}{\alpha}}}
 {}_2F_1\left(1,\frac{\sqrt{6}-3\alpha}{\sqrt{6}-2\alpha},\frac{\sqrt{6}-4\alpha}{\sqrt{6}-2\alpha};-\frac{
 z^{2-\frac{\sqrt{6}}{\alpha}}}{\epsilon \mu^2/p_{\bar \phi}^2}\right) \,.
\end{align}
Thus we obtain
\begin{align}
    a =& \Bigg\{ \frac{\alpha^2p_{\bar\phi}^2}{12\sin(2\beta)}
    \left(\frac{6}{\alpha^2}(F^{-1}(\tilt-\tilt_0))^2-(\tilt-\tilt_0+\bar{t}_0)^2 \right) 
    \Bigg\}^{\frac{1}{2\sqrt{6}\alpha}} \,, \\
    \phi =& \frac{\sqrt{6}}{3}\ln \left[-\tan(\beta)
    \frac{(\tilt-\tilt_0+\bar{t}_0)+(\sqrt{6}/\alpha)F^{-1}(\tilt-\tilt_0)}
    {(\tilt-\tilt_0+\bar{t}_0)-(\sqrt{6}/\alpha)F^{-1}(\tilt-\tilt_0)}\right] \,.
\end{align}
Finally, we convert the time $\tilt$ to $t$ to obtain
the solutions to the equations of motion for $H$ by 
\begin{align}
 t = \int\Omega^{-1+\frac{\sqrt{6}}{4\alpha}}d\tilt = \int
 \left(\frac{\alpha^2p_{\bar\phi}^2}{12\sin(2\beta)}
    \left(\frac{6}{\alpha^2}(F^{-1}(\tilt-\tilt_0))^2-(\tilt-\tilt_0+\bar{t}_0)^2 \right)\right)^{-1+\frac{\sqrt{6}}{4\alpha}}d\tilt \,.
\end{align}

\subsection{Solutions: Case IV}
\label{appendixB-4}

This case is the general case when the field space becomes conformally flat. 
Still in this case, since we have $[\xi_{(1)}, \xi_{(3)}]=0$, we can introduce new coordinates 
$( \bar{a}(a,\phi), \bar{\phi}(a,\phi), \chi)$
in which  $\xi_{(3)}$ is written as
$\xi_{(3)}=\partial/\partial \bar{\phi}$.
Actually, introducing the coordinate transformation
\begin{align}
 \bar{a} =& a^{\sqrt{6}\alpha}e^{\mp\alpha\phi} \,, \\
 \bar{\phi} =& \pm\frac{1}{2\sqrt{6}\alpha} a^{\sqrt{6}\alpha} e^{\pm\alpha\phi} \,,
\end{align}
or, equivalently,
\begin{align}
 a =& \left(\pm
 2\sqrt{6}\alpha \bar{a}\bar{\phi}\right)^{\frac{1}{2\sqrt{6}\alpha}} \,, \\
 \phi =& \pm\frac{1}{2\alpha}
 \ln \left[
 \pm\frac{ 2\sqrt{6}\alpha\bar{\phi}}{\bar{a}}
 \right] \,,
\end{align}
the Hamiltonian is given by
\begin{align}
    H = \Omega^{1-\frac{\sqrt{6}}{4\alpha}} H_0 \,,
\end{align}
where
\begin{align}
 H_0 =& \mp\frac{\alpha}{\sqrt{6}}p_{\bar a} p_{\bar \phi}
 +  \bar{a}^{-2+\frac{\sqrt{6}}{\alpha}}V_0p_\chi^2  \,, \\
 \Omega =& \pm 2\sqrt{6}\alpha \bar{a}\bar{\phi} \,.
 \label{hamiltonian-case4}
\end{align}
As shown in the previous subsection, from the conformal invariance of 
null geodesics, the solutions for the Hamiltonian $H$ are obtained  from the 
solutions for the Hamiltonian $H_0$ by changing the time coordinate. 
The equations of motion for $H_0$ are given by
\begin{align}
 \dot{\bar{a}} =& \mp\frac{\alpha}{\sqrt{6}} p_{\bar \phi} \,,
 \quad \dot{p}_{\bar a} =-\left(-2+\frac{\sqrt{6}}{\alpha}\right)  
 \bar{a}^{-3+\frac{\sqrt{6}}{\alpha}} V_0p_\chi^2 \,,
 \label{eom-41}\\
 \dot{\bar{\phi}} =& \mp\frac{\alpha}{\sqrt{6}}  p_{\bar a} \,,
 \quad \dot{p}_{\bar \phi} = 0 \,, 
 \label{eom-42}\\
 \dot{\chi} =& 2 \bar{a}^{-2+\frac{\sqrt{6}}{\alpha}}V_0  p_\chi \,,
 \quad \dot{p}_\chi = 0 \,.
 \label{eom-43}
\end{align}
Eq.~(\ref{eom-41}) can be easily solved by
\begin{align}
    &\bar{a} =
    \mp\frac{\alpha}{\sqrt{6}}p_{\bar\phi}(\tilt-\tilt_0) \,,
\end{align}
and  for $\alpha\neq \sqrt{6}$ Eqs.~({\ref{eom-42})-and (\ref{eom-43})  can be solved by
\begin{align} 
    \bar{\phi} =&
    -\frac{1}{-1+\frac{\sqrt{6}}{\alpha}}\frac{V_0p_\chi^2}{p_{\bar\phi}}
    \left(\mp \frac{\alpha}{\sqrt{6}}p_{\bar\phi}\right)^{-2+\frac{\sqrt{6}}{\alpha}}
    (\tilt-\tilt_0)^{-1+\frac{\sqrt{6}}{\alpha}}+\bar{\phi}_0 \,,\\
    \chi =& 
    \frac{2}{-1+\frac{\sqrt{6}}{\alpha}}
    V_0 p_\chi \left(\mp \frac{\alpha}{\sqrt{6}}p_{\bar\phi}\right)^{-2+\frac{\sqrt{6}}{\alpha}}
    (\tilt-\tilt_0)^{-1+\frac{\sqrt{6}}{\alpha}}
    +\chi_0 \,,
\end{align}
while for $\alpha=\sqrt{6}$ 
\begin{align}
  \bar{\phi} =&
    \pm\frac{V_0p_{\chi}^2}{p_{\bar\phi}^2}\ln p_{\bar\phi}(\tilt-\tilt_0)+\bar{\phi}_0 \,,\\
    \chi =& \mp\frac{2V_0p_{\chi}}{p_{\bar\phi}}\ln p_{\bar\phi}(\tilt-\tilt_0) 
    +\chi_0 \,.
\end{align}
Thus the solutions to Eqs.~\eqref{Friedmann-Eq-1}--\eqref{Friedmann-Eq-3} are given by for $\alpha\neq \sqrt{6}$
\begin{align}
 a =& \Bigg\{
 \frac{2\alpha^2V_0p_\chi^2}{-1+\frac{\sqrt{6}}{\alpha}}
 \left(\mp \frac{\alpha p_{\bar\phi}}{\sqrt{6}}\right)^{-2+\frac{\sqrt{6}}{\alpha}}(\tilt-\tilt_0)^{\frac{\sqrt{6}}{\alpha}}
 -2\alpha^2\bar{\phi}_0p_{\bar\phi}(\tilt-\tilt_0)\Bigg\}^{\frac{1}{2\sqrt{6}\alpha}} \,,\label{eq:IV72}\\
 \phi =& \pm\frac{1}{2\alpha}
 \ln\left[
 \frac{12V_0p_\chi^2}{\left(-1+\frac{\sqrt{6}}{\alpha}\right)p_{\bar\phi}^2}
 \left(\mp \frac{\alpha p_{\bar\phi}}{\sqrt{6}}\right)^{-2+\frac{\sqrt{6}}{\alpha}}
 (\tilt-\tilt_0)^{-2+\frac{\sqrt{6}}{\alpha}}
 -\frac{12\bar{\phi}_0}{p_{\bar\phi}(\tilt-\tilt_0)} 
 \right]\,,\label{eq:IV73}
\end{align}
with the reparameterization of the time coordinate from $\tilt$ to $t$ by
\begin{align}
 t =\int \Bigg\{
 \frac{2\alpha^2V_0p_\chi^2}{-1+\frac{\sqrt{6}}{\alpha}}
 \left(\mp \frac{\alpha p_{\bar\phi}}{\sqrt{6}}\right)^{-2+\frac{\sqrt{6}}{\alpha}}(\tilt-\tilt_0)^{\frac{\sqrt{6}}{\alpha}}
 -2\alpha^2\bar{\phi}_0p_{\bar\phi}(\tilt-\tilt_0)\Bigg\}^{-1+\frac{\sqrt{6}}{4\alpha}} d\tilt \,.
 \label{eq:IV74}
\end{align}
For $\alpha=\sqrt{6}$, the solutions are given by
\begin{align}
a=&\Bigg\{\mp\frac{12V_0p_{\chi}^2}{p_{\bar\phi}}(\tilt-\tilt_0)\ln p_{\bar\phi}(\tilt-\tilt_0)-12\bar{\phi}_0p_{\bar\phi}(\tilt-\tilt_0)
\Bigg\}^{\frac{1}{12}}\,,\\
\phi=&\pm\frac{1}{2\sqrt{6}}\ln\left[\mp\frac{12V_0p_{\chi}^2}{p_{\bar\phi}^3}\frac{\ln p_{\bar\phi}(\tilt-\tilt_0)}{\tilt-\tilt_0}-\frac{12\bar{\phi}_0}{p_{\bar\phi}(\tilt-\tilt_0)}\right]
\,,
\end{align}
with
\beqa
t=\int \Bigg\{
\mp\frac{12V_0p_{\chi}^2}{p_{\bar\phi}}(\tilt-\tilt_0)\ln p_{\bar\phi}(\tilt-\tilt_0)-12\bar{\phi}_0p_{\bar\phi}(\tilt-\tilt_0)
\Bigg\}^{-\frac34}d\tilt\,. 
\eeqa

For example, by setting $\bar{\phi}_0=0$, from Eqs.~(\ref{eq:IV72}) and (\ref{eq:IV73}), 
we have $a\propto (\tilt-\tilt_0)^{\frac{1}{2\alpha^2}}$ 
 and $\phi=\pm\frac{\sqrt{6}-2\alpha}{2\alpha^2}\ln (\tilt-\tilt_0)+{\rm const.}$. 
 Moreover, from Eq. (\ref{eq:IV74}), $t$ is given by 
 $t-t_0\propto (\tilt-\tilt_0)^{\frac{(\sqrt{6}-2\alpha)^2}{4\alpha^2}}$ 
 where $t_0$ is a constant. 
We  thus obtain 
\beqa
a&\propto & (t-t_0)^{\frac{2}{(\sqrt{6}-2\alpha)^2}},\\
\phi&=&\pm \frac{2}{\sqrt{6}-2\alpha}\ln (t-t_0)+{\rm const.}\,, 
\eeqa
reproducing the well-known result for power-law inflation 
\cite{Lucchin:1984yf,Chiba:2024iia}.

\section{Discussion}
\label{sec6}


We have so far only considered time-dependent system. 
Then, what about a spacetime-dependent system?
Ref. \cite{Finn:2018cfs} suggested  introducing a vierbein $e^{\mu}_M$ satisfying 
$\nabla_\mu e^\nu_M=0$ in order to expand  the vector field as
\beqa
B^{\mu}=\phi^Me^{\mu}_M\,,
\eeqa
where $\phi^M~ (M=n+1,\dots,n+4)$ are four scalar fields.
Then the Lagrangian of the scalar fields and the vector field 
in Eq. (\ref{action:scalar2}) can be 
written as  purely kinetic terms of $n+4$ scalar fields
\beqa
{\cal L}_R=-\frac12 \sqrt{-g}g^{\mu\nu}k_{IJ}\p_{\mu}\phi^I\p_{\nu}\phi^J
+\frac{1}{4V}\sqrt{-g}e^{\mu}_{(M}e^{\nu}_{N)}\p_{\mu}\phi^M\p_{\nu}\phi^N
\equiv -\frac12\sqrt{-g}H^{\mu\nu}_{AB}\p_{\mu}\phi^A\p_{\nu}\phi^B\,,
\eeqa
where $\phi^A (A=1,\dots,n+4)$ are $n+4$ scalar fields and 
$H^{\mu\nu}_{AB}$ is an extended metric on a mixed space of 
a spacetime and a field space and is given by
\beqa
H^{\mu\nu}_{AB}=
\begin{pmatrix}
g^{\mu\nu}k_{IJ} & 0 \\
0 & -\frac{1}{2V}e^{\mu}_Me^{\nu}_N \\
\end{pmatrix}
.
\eeqa
The equation of motion of $\phi^A$ is then written as
\beqa
H^{\mu\nu}_{AB}\nabla_{\mu}\p_{\nu}\phi^B+H^{\mu\nu}_{AD}\Gamma^{D}_{BC}\p_{\mu}\phi^B\p_{\nu}\phi^C=0\,,
\label{canonical2-1}
\eeqa
where $\Gamma^D_{BC}$ is the connection defined with respect to $H^{\mu\nu}_{AB}$ via $\nabla_AH^{\mu\nu}_{BC}=0$ and is given by
\beqa
H^{\mu\nu}_{AD}\Gamma^D_{BC}&=&\frac12\left(\p_BH^{\mu\nu}_{CA}+\p_CH^{\mu\nu}_{BA}-\p_AH^{\mu\nu}_{BC}\right)\,.
\label{connection:AB}
\eeqa

Then, it is tempting to search for the KVss and KTs of the extended field space metric $H^{\mu\nu}_{AB}$ by solving the Killing equations, 
such as $\nabla_{(A}\xi_{B)}=0$ or $\nabla_{(A}K_{BC)}=0$. 
In order to do that, it is necessary to compute $\Gamma^A_{BC}$ 
by solving Eq. (\ref{connection:AB}) for $\Gamma^D_{BC}$. 
However,  Eq. (\ref{connection:AB}) are overdetermined system for $\Gamma^D_{BC}$:  Eq. (\ref{connection:AB}) are ten equations for 
each $\Gamma^D_{BC}$ with ten components of  $\mu,\nu$, and 
$\Gamma^D_{BC}$ is not determined in general. 
Therefore,  the extended field space metric $H^{\mu\nu}_{AB}$ may not be 
helpful in studying the symmetry of the system.

\section{Summary}
\label{sec7}

We have applied the Eisenhart lift  of Riemannian type  to 
the system of a scalar field in a flat FLRW universe and studied 
the condition for the existence of constants of motion so that 
the system is integrable.  
We have found that  KV fields and a KT field  exist 
 for a particular shape of potential  
(\ref{potential-I}) extending the previous results in the literature. 
 Moreover, we have found that a CKV field exists 
for a potential written as an exponentiation of a combination of exponential potentials with general index 
(\ref{potential-III}).  In particular,  for a simple exponential potential 
with general index which corresponds to the potential   for power-law 
inflation (\ref{potential-IV}),  we have found that the field space is 
conformally flat and have obtained explicitly the maximal number of CKVs. 
We have solved the equations of motion by introducing a coordinate along the (C)KV and given the complete solutions to the equations of motion. 
We have also studied the condition for the existence of CKVs 
for the system of multiple scalar fields. 
It would be interesting to study quantum cosmology for these integrable systems 
generalizing the results in \cite{Garay:1990re}.

We have  pointed out the problem in extending the Eisenhart lift to a 
spacetime-dependent system. Also, we have attempted to extend 
the Eisenhart lift of Lorentzian type to scalar field theories by 
introducing two vector fields and have found the problem in finding 
the symmetry of the system. 

It would be interesting to extend the analysis to other homogeneous 
cosmological spacetime. 
It would also be interesting to classify 
the integrable system with the kinetic term of Lorentzian signature. 
We hope to report the results of the investigation in the future.

\section*{Acknowledgments}
This work is supported by JSPS Grant-in-Aid for Scientific Research Number 
22K03640  (TC) and in part by Nihon University. 


\appendix

\section{Central Force Problem}
\label{appendixA}

In this Appendix, as an example of the use of a lifted system, 
we consider a particle moving about a fixed center of force. 
Taking the center of force as the origin of the coordinate,    
the potential is a function of $r=|\x|$, $V(r)$. 

\subsection{Riemannian Lift}

The lifted Lagrangian is given by
\beqa
L_R=\frac12\dot\x^2
+\frac{1}{4V(r)}\dot y^2=\frac12 G_{AB}\dot x^A\dot x^B\,,
\eeqa
where $x^A=(x^i,y)$ with $i=1,\dots,3$ and $G_{ij}=\delta_{ij},G_{iy}=0$ and 
$G_{yy}=\frac{1}{2V(r)}$.  
In addition to the usual three KVs associated 
with the conservation of angular momentum ${\bL}={\x}\times {\bp}$, 
there may exist  yet another constant of motion associated with 
a KT $K^{AB}$ which satisfies $\nabla_{(A}K_{BC)}=0$,  
where the bracket in the index denotes the symmetrization.  
{} From the KT equations, we have
\beqa
&&\p_iK_{jk}+\p_jK_{ki}+\p_kK_{ij}=0,\label{killing1}\\
&&\p_iK_{jy}+\p_jK_{iy}+\frac{V'}{rV}\left(x_iK_{jy}+x_jK_{iy}\right)=0,\label{killing2}\\
&&x^iK_{iy}=0,\label{killing3}\\
&&\p_iK_{yy}+\frac{2x_iV'}{rV}K_{yy}-\frac{V'}{2rV^2}K_{ij}x^j=0
\label{killing4}\,,
\eeqa
where $V'=dV/dr$. The most general solution to Eqs. (\ref{killing2}) and (\ref{killing3}) is $K_{iy}=\frac{1}{V}\epsilon_{ijk}a^jx^k$ for a constant vector $a^i$. 
Also, the most general solution to Eq. (\ref{killing1})   
 is  $K_{ij}=b_{(i}x_{j)}-(\bb\cdot\x)\delta_{ij}+C_{ij}$ for 
a constant vector $b_i$ and a constant tensor $C_{ij}$. 
Plunging these into Eq. (\ref{killing4}), the integrability condition of 
$K_{yy}$  gives \cite{1984RpMP...20...31C}
\beqa
\x\times \left(\left(r\left(\frac{V'}{r}\right)'+3\frac{V'}{r}\right)\bb+\frac{1}{r}\left(\frac{V'}{r}\right)'\bC\right)=0\,,
\label{integrability}
\eeqa
where we have introduced $C_i=C_{ij}x^j$. 

Excluding the trivial case where $V$ is constant, 
the condition Eq. (\ref{integrability}) is satisfied if 
(i) $r\left(\frac{V'}{r}\right)'+3\frac{V'}{r}=0$ and $\bC\propto \x$, or 
(ii) $\left(\frac{V'}{r}\right)'=0$ and $\bb=0$. 
The former case implies $V= -\frac{\mu}{r}$ 
with $\mu$ being a constant (Newtonian potential) and $C_{ij}=C\delta_{ij}$ with $C$ being a constant, while 
the latter case implies $V=\frac12 kr^2$ with $k$ being a constant (harmonic potential). 

For case (i), $V= -\frac{\mu}{r}$, the solution to Eq. (\ref{killing4}) is  
$K_{yy}=D\frac{r^2}{4\mu^2}+\frac{r}{4\mu}\left((\bb\cdot\x)-2C\right)$, where 
$D$ is a constant. 
Then, the corresponding constant of motion is
\beqa
K^{AB}p_Ap_B&=&
(\bb\cdot\bp)(\x\cdot\bp)-(\bb\cdot\x){\bp}^2 +\frac{\mu}{r}(\bb\cdot\x)p_y^2
+2C\left(\frac{\bp^2}{2}-\frac{\mu}{r}p_y^2\right)+4p_y \left(\ba\cdot \bL\right)+Dp_y^2\nonumber\\
&=&-\bb\cdot\left(\bp\times (\x\times \bp)-\frac{\mu}{r}\x\right)+
2C\left(\frac{\bp^2}{2}-\frac{\mu}{r}\right) + 4 \ba\cdot \bL+D
\label{conserved:k-tensor}
\,.
\eeqa
where we have set $p_y=1$ in the second line.  
The last three terms in Eq. (\ref{conserved:k-tensor}) are the Hamiltonian, the $\ba$ component of the angular momentum and a constant, respectively. 
The first term in Eq. (\ref{conserved:k-tensor}) is nothing but the 
$\bb$ component of the Laplace-Runge-Lenz vector \cite{goldstein,1984RpMP...20...31C}
\beqa
\A={\bp} \times {\bL}-\frac{\mu}{r}\x\,.
\label{lenz:vector}
\eeqa

For case (ii),  $V=\frac12 kr^2$, the solution to Eq. (\ref{killing4}) is 
$K_{yy}=\frac{D}{k^2r^4}+\frac{1}{kr^4}C_{ij}x^ix^j$, where $D$ is a constant. 
The constant of motion is thus
\beqa
K^{AB}p_Ap_B&=&C^{ij}\left(p_ip_j+kx_ix_jp_y^2\right)+4p_y \left(\ba\cdot \bL\right)+Dp_y^2\nonumber\\
&=&C^{ij}\left(p_ip_j+kx_ix_j\right)+4 \ba\cdot \bL+D\,,
\eeqa
where we have set $p_y=1$ in the second line.  
Since $C_{ij}$ is an arbitrary constant tensor, the existence 
of the constant of motion implies the conservation of 
the Laplace-Runge-Lenz tensor \cite{goldstein,1984RpMP...20...31C}
\beqa
A_{ij}=p_ip_j+k x_ix_j\,.
\label{lenz:tensor}
\eeqa

To summarize,  the only nontrivial potential which admits KT 
is either Newtonian potential or harmonic potential, and the corresponding 
constant of motion is the Laplace-Runge-Lenz vector for the former,   
the Laplace-Runge-Lenz tensor for the latter.

\subsection{Coordinate Transformation}

It is interesting to note that the system with Newtonian potential is related to the system with harmonic potential by a simple coordinate transformation. 

Taking the plane of the motion in $xy-$plane, the Lagrangian of the Kepler problem is given by
\beqa
L=\frac12 (\dot x^2+\dot y^2)+\frac{\mu}{\sqrt{x^2+y^2}}\,.
\eeqa
We introduce new coordinates $x'$ and $y'$ by the following coordinate transformation
\beqa
x+iy=\frac12(x'^2-y'^2)+ix'y'\,.
\eeqa
Then $L$ becomes
\beqa
L=\frac12 (x'^2+y'^2)(\dot x'^2+\dot y'^2)+\frac{2\mu}{x'^2+y'^2}\,,
\eeqa 
and in terms of conjugate momenta $p_{x'}=(x'^2+y'^2)\dot x',p_{y'}=
(x'^2+y'^2)\dot y'$ the Hamiltonian is given by
\beqa
H=\frac{1}{2(x'^2+y'^2)}(p_{x'}^2+p_{y'}^2)-\frac{2\mu}{x'^2+y'^2}\,.
\label{H-kepler-harmonic}
\eeqa
Since we consider a bound orbit, $H$ is a negative constant and we set $H=-\frac{\omega^2}{2}$ with $\omega$ being a constant. 
Then, Eq. (\ref{H-kepler-harmonic}) can be rewritten as
\beqa
\frac12(p_{x'}^2+p_{y'}^2)+\frac12 \omega^2(x'^2+y'^2)=2\mu\,,
\eeqa
which is nothing but the Hamiltonian for the harmonic potential.

\section{Integrability Conditions of CKV Equation}
\label{appendixC}

In this appendix, we summarize the bound on the dimension of the vector space of CKV fields and the integrability conditions of the CKV equation (CKVE).

\subsection{Bound on the Dimensionality}
For a vector field $\xi^\mu$ on an $n$-dimensional spacetime $(M,g_{\mu\nu})$,
the CKVE is given by
\beqa
 \nabla_{(\mu}\xi_{\nu)} = Q g_{\mu \nu} \,,
 \label{CKV_equation}
\eeqa
where $\nabla_\mu$ is the covariant derivative with 
the Levi-Civita connection of a metric $g_{ab}$.
From Eq.~\eqref{CKV_equation}, it is easy to show that
\beqa
 Q = \frac{1}{n} \nabla_\mu\xi^\mu \,. \label{Def_associated_func}
\eeqa
This function $Q$ is called the associated function of $\xi^\mu$.
Since the CKVE is linear, the space of CKVs is a vector space. In two dimensions, this vector space of CKVs is of infinite dimensions. On the other hand, in $n\geq 3$ dimensions, the vector space is of finite dimensions since the CKVE becomes an overdetermined system of first-order linear PDEs:
\beqa
 \nabla_\mu \xi_\nu
&=& L_{\mu\nu} + g_{\mu\nu} Q \,, \label{prolonged_eq21}\\
 \nabla_\mu L_{\nu\rho}
&=& - W_{\nu\rho\mu}{}^\sigma\xi_\sigma
    - 2S_{\mu[\nu}\xi_{\rho]}-2g_{\mu[\nu}\eta_{\rho]} \,, \label{prolonged_eq22}\\
 \nabla_\mu Q
&=& - S_\mu{}^\nu\xi_\nu + \eta_\mu \,, \label{prolonged_eq23}\\
 \nabla_\mu \eta_\nu
&=& - C_{\nu\mu}{}^\rho \xi_\rho  - S^\rho{}_\mu L_{\nu\rho}
    - S_{\mu\nu} Q \,, \label{prolonged_eq24}
\eeqa
where
\beqa
 L_{\mu\nu} \equiv \nabla_{[\mu}\xi_{\nu]} \,, \quad
 \eta_\mu \equiv \partial_\mu Q
 + S_\mu{}^\nu\xi_\nu \,, \label{Def_eta}
\eeqa
and $W_{\mu\nu\rho}{}^\sigma$, $S_{\mu\nu}$ and $C_{\mu\nu\rho}$ 
are the Weyl tensor, Schouten tensor and Cotton tensor, respectively.  
Schouten tensor and Cotton tensor are defined by
\beqa
 S_{\mu\nu} &=& \frac{1}{n-2}\left(R_{\mu\nu}
 -\frac{R}{2(n-1)}g_{\mu\nu}\right) \,,
 \label{def_Schouten}\\
  C_{\mu\nu\rho}
&=& \nabla_\rho S_{\mu\nu} - \nabla_\nu S_{\mu\rho} \,.
\label{Cotton-York}
\eeqa
We find from Eqs.~\eqref{prolonged_eq21}--\eqref{prolonged_eq24} that if the values of the variables
$(\xi_\mu,L_{\mu\nu},Q,\eta_\mu)$ are given at a point on $M$, 
the values of these variables are determined on the entire $M$. 
Since the number of initial values to provide is given by
\beqa
 n + \frac{n(n-1)}{2} + 1 + n = \frac{(n+1)(n+2)}{2} \,,
\eeqa
the number of linearly independent solutions to the PDEs (i.e. CKVs) is bounded by this number.

\subsection{Integrability conditions}
Further bounds to the dimension of the vector space of CKVs are obtained by the integrability conditions. To obtain the integrability conditions, we regard the set of variables ${\cal Z}_A \equiv (\xi_\mu, L_{\mu\nu}, Q,\eta_\mu)$
as a section of the vector bundle $E=T^*M\oplus \Lambda^2T^*M\oplus C^\infty(M) \oplus T^*M$,
and regard Eqs.~\eqref{prolonged_eq21}--\eqref{prolonged_eq24} as the parallel section equation on $E$,
\beqa
 D_\mu {\cal Z}_A = 0 \,,
\eeqa
with respect to the Killing connection $D_\mu$
defined by
\beqa
 D_\mu {\cal Z}_A
\equiv
\left(
\begin{array}{c}
 \nabla_\mu \xi_\nu - L_{\mu\nu} - g_{\mu\nu} Q \\
 \nabla_\mu L_{\nu\rho}
 + W_{\nu\rho\mu}{}^\sigma\xi_\sigma
    + 2S_{\mu[\nu}\xi_{\rho]} + 2g_{\mu[\nu}\eta_{\rho]} \\
 \nabla_\mu Q
  + S_\mu{}^\nu\xi_\nu - \eta_\mu \\
 \nabla_\mu \eta_\nu
  + C_{\nu\mu}{}^\rho \xi_\rho  + S^\rho{}_\mu L_{\nu\rho}
    + S_{\mu\nu} Q
\end{array}
\right) \,.
 \label{Prolongation_Connection_2}
\eeqa
Then the integrability conditions are obtained by computing the curvature of $D$,
\beqa
 R^D_{\mu\nu A}{}^B{\cal Z}_B \equiv
 (D_\mu D_\nu - D_\nu D_\mu){\cal Z}_A= 0 \,.
\eeqa
These equations are an overdetermined system of algebraic equations to ${\cal Z}_A$ so that the number of the solutions to these equations provides the bound to the dimension of the vector space of CKVs.

\paragraph{$n \geq 4$ dimensions.}
From a direct calculation, we find that
$(D_\mu D_\nu - D_\nu D_\mu)\xi_\rho = 0$
 nor $(D_\mu D_\nu - D_\nu D_\mu)Q = 0$
 do not provide nontrivial conditions;
 on the other hand, $(D_\mu D_\nu - D_\nu D_\mu)L_{\rho\sigma} = 0$
 gives rise to the first integrability conditions
\begin{align}
 &\nabla_{[\mu}W_{|\rho\sigma|\nu]}{}^\lambda \xi_\lambda
 -C_{[\rho|\mu\nu|}\xi_{\sigma]}
 +2 g_{[\mu|[\rho}C_{\sigma]|\nu]}{}^\lambda \xi_\lambda \nonumber\\
&  +W_{\mu\nu[\rho}{}^\lambda L_{|\lambda|\sigma]}
  +W_{\rho\sigma[\mu}{}^\lambda L_{|\lambda|\nu]}
  -W_{\mu\nu\rho\sigma} Q =0 \,,
  \label{int_cond_1}
\end{align}
and $(D_\mu D_\nu - D_\nu D_\mu)\eta_\rho = 0$ gives rise to
the second integrability conditions
\begin{align}
 (\nabla^\lambda C_{\mu\nu\rho} + W_{\nu\rho\mu}{}^\lambda)\xi_\lambda
 +2C_{\mu[\nu}{}^\lambda L_{\rho]\lambda}
  +C^\lambda{}_{\nu\rho} L_{\mu\lambda}
 +3C_{\mu\nu\rho} Q -W_{\nu\rho\mu}{}^\lambda \eta_\lambda = 0 \,.
  \label{int_cond_2}
\end{align}

It should be noted that the integrability conditions
are written in terms of Weyl tensor.
If a spacetime is conformally flat, the Weyl tensor vanishes
and the integrability conditions are automatically satisfied.
This means that we obtain the maximum number of CKVs on a conformally flat spacetime (see, e.g. \cite{Batista:2017wcm}).

\paragraph{3 dimensions.}
Because of the speciality of 3 dimensions,
the first conditions \eqref{int_cond_1} 
are automatically satisfied
and the second conditions \eqref{int_cond_2} reduce to
\begin{align}
 (\nabla^\lambda C_{\mu\nu\rho})\xi_\lambda
 +2 C_{\mu[\nu}{}^\lambda L_{\rho]\lambda}
 +C^\lambda{}_{\nu\rho} L_{\mu\lambda}
 +3C_{\mu\nu\rho} Q  = 0 \,.
 \label{3D_int_cond_1}
\end{align}
Moreover, taking the covariant derivative of
\eqref{3D_int_cond_1}, we obtain further conditions
\begin{align}
& (\nabla^\lambda \nabla_\sigma C_{\mu\nu\rho})\xi_\lambda
 -3C_{\mu\nu\rho}\overline{\xi}_\sigma \nonumber\\
& -C_{\sigma\nu\rho}\overline{\xi}_\mu
 +C^\lambda{}_{\nu\rho} g_{\sigma\mu}\overline{\xi}_\lambda
 +2 C_{\mu\sigma[\nu} \overline{\xi}_{\rho]}
 -2 C_\mu{}^\lambda{}_{[\nu} g_{\rho]\sigma}\overline{\xi}_\lambda \nonumber\\
& + (\nabla^\lambda C_{\mu\nu\rho})L_{\sigma\lambda}
  + 2(\nabla_\sigma C_{\mu[\nu}{}^\lambda)L_{\rho]\lambda}
  + (\nabla_\sigma C^\lambda{}_{\nu\rho})L_{\mu\lambda} \nonumber\\
& - 2C_{\mu\nu}{}^\lambda g_{\sigma[\rho}\eta_{\lambda]}
  + 2C_{\mu\rho}{}^\lambda g_{\sigma[\nu}\eta_{\lambda]}
  - 2C^\lambda{}_{\nu\rho} g_{\sigma[\mu}\eta_{\lambda]}
  + 3C_{\mu\nu\rho}\eta_\sigma =0 \,,
\label{3D_int_cond_2}
\end{align}
where
\beqa
 \overline{\xi}_\mu = S_\mu{}^\nu \xi_\nu \,.
\eeqa

When applied to the field space metric Eq. (\ref{effective-metric}) with a potential $V(\phi)$ arbitrary, Eq.~(\ref{3D_int_cond_1}) and Eq.~(\ref{3D_int_cond_2}) fix eight components  
 out of ten components of $(\xi_\mu, L_{\mu\nu}, Q,\eta_\mu)$,  
 which implies that the number of independent CKVs is two.

\section{Symmetires in Multiple Scalar Field System}
\label{multi}

In this Appendix, we consider the condition for the existence of 
CKVs for multiple scalar fields. 
For simplicity, we study two scalar fields. 
The Lagrangian of  the scalars in a flat FLRW universe is
\beqa
L=\frac12\left(-6a\dot a^2+a^3\left(\dot\phi_1^2+\dot\phi_2^2\right)\right)-
a^3V(\phi_1,\phi_2)\,.
\label{lagrangian}
\eeqa
The system Eq. (\ref{lagrangian}) is lifted by introducing one vector field $B^{\mu}$:
\beqa
L_R=\frac12\left(-6a\dot a^2+a^3\left(\dot\phi_1^2+\dot\phi_2^2\right)\right)
+\frac{1}{4a^3}\frac{\dot\chi^2}{V}\,,
\eeqa 
where $\chi=a^3B^0$.

\subsection{Case A: Separable form}  

First, we consider the potential of separable  form $V(\phi_1,\phi_2)=V_1(\phi_1)+V_2(\phi_2)$. 

In this case, we find that CKV exists for two kinds of potential. 

\subsubsection{Case A-1}
The first case is an exponential potential: 
\beqa
V(\phi_1,\phi_2)=c_1e^{\lambda_1\phi_1}+c_2
e^{\lambda_2\phi_2}\,,
\label{multi:a1}
\eeqa
where $c_1,c_2,\lambda_1$ and $\lambda_2$ are constants. 
By shifting the scalar field by a constant, one can always set $c_1=c_2$ 
without loss of generality.  
For this potential,  
 we find a CKV exists such that
\beqa
\xi_{(1)}=-\frac{a}{6}\frac{\p}{\p a}+
\frac{1}{\lambda_1}\frac{\p}{\p \phi_1}+\frac{1}{\lambda_2}\frac{\p}{\p \phi_2},
\eeqa
The potential (\ref{multi:a1}) corresponds to the potential for assisted inflation \cite{Liddle:1998jc,Malik:1998gy} where inflation can proceed 
even if each potential is too steep to sustain inflation. 
However, we do not find another independent CKV. 

It is to be noted that although it has been shown \cite{Liddle:1998jc} 
that the system is equivalent to 
 a single scalar field  system for a scaling solution,  it does not imply that the degrees of freedom of the system is one dimensional so that the system is completely integrable because there exist degrees of freedom orthogonal to 
 the scaling solution. 


\subsubsection{Case A-2}
The second case is a polynomial potential of the form:
\beqa
V(\phi_1,\phi_2)=V_0\left(\phi_1^2+\phi_2^2+a_1\phi_1+a_2\phi_2+b\right)\,,
\eeqa
where $V_0,a_1,a_2$ and $b$ are constants.
For this potential, we find a KV such that
\beqa
\xi_{(1)}=(2\phi_2+a_2)\frac{\p}{\p \phi_1}-(2\phi_1+a_1)\frac{\p}{\p \phi_2}\,.
\eeqa
We do not find another  independent  CKVs. 

\subsection{Case B: Factorized form}
 Secondly, we consider the potential of factorized  form $V(\phi_1,\phi_2)=V_1(\phi_1)V_2(\phi_2)$.  

In this case, 
we find two CKVs exist for the potential of the form
\beqa 
V=V_0e^{\lambda(\phi_1+\phi_2)}\,,
\label{pot:b}
\eeqa
where $V_0$ and  $\lambda$ are constants, 
\beqa
\xi_{(1)}&=&\frac{\p}{\p\phi_1}-\frac{\p}{\p\phi_2}\,,\\
\xi_{(2)}&=&-\frac{f}{6}a\frac{\p}{\p a}+g\frac{\p}{\p\phi_1}+\left(\frac{f}{\lambda}-g\right)\frac{\p}{\p\phi_2}\,,
\eeqa
where $f$ and $g$ are constants. $\xi_{(1)}$ is a KV and $\xi_{(2)}$ is a proper CKV. 
The potential (\ref{pot:b}) has been proposed in \cite{Kanti:1999vt,Copeland:1999cs} 
as a generalization of assisted inflation. 
Since, these two CKVs commute, $[\xi_{(1)},\xi_{(2)}]=0$, there exist two independent constants of motion. Therefore, 
together with the Hamiltonian constraint and a constant of motion 
associated with $\p/\p\chi$, we have four constants of motion and 
hence the system is completely integrable.

\section{Einsenhart Lift of Lorentzian type}
\label{sec4}

In this appendix, after reviewing the Eisenhart lift of Lorentzian type 
for a particle, we attempt to construct the Einsenhart lift of Lorentzian 
type for scalar fields. 

\subsection{Eisenhart Lift for a Particle}

Eisenhart showed that the equation of motion of a particle Eq. (\ref{eom1}) 
can also be derived from the action by adding two new coordinates $u$ and $v$
\cite{eisenhart1928,Cariglia:2015bla}
\beqa
I_L=\int dt\left(\frac12 \delta_{ij}\dot x^i\dot x^j+\dot u\dot v-V(x^1,\dots,x^n)\dot u^2\right) \equiv \int dt~\frac12 G_{AB}\dot x^A\dot x^B\,,
\label{action3}
\eeqa
where we have introduced $n+2$ dimensional coordinate $x^A=(x^i,u,v)$ and 
$G_{ij}=\delta_{ij},G_{uu}=-2V,G_{uv}=1$. It is a so-called Lorentzian lift since the signature of the field space metric is negative (if $V>0$). 
Note that, unlike the case of Riemannian lift, the metric is always well defined for any $V(\x)$. 

The equation of $x^i$ and $u,v$ are
\beqa
\ddot x^i&=&-\left(\p_iV\right)\dot u^2\,,
\label{eom3-1}\\
p_u&=&\dot v-2V\dot u={\rm const}\,,
\label{eom3-2}\\
p_v&=&\dot u={\rm const}\,.
\label{eom3-3}
\eeqa
Then by setting $p_v=1$ reproduces Eq. (\ref{eom1}). Moreover, 
if we restrict the solutions of the equation of motion to those which are 
null geodesics of Eq. (\ref{action3})
\beqa
0=\frac12 \delta_{ij}\dot x^i\dot x^j+\dot u\dot v-V\dot u^2=\frac12 \delta_{ij}\dot x^i\dot x^j +
\dot v-V,
\eeqa
then $p_u$ coincides with the minus of the Hamiltonian 
$H$ of the action  
Eq. (\ref{action1}):
\beqa
p_u=\dot v-2V=-\frac12 \delta_{ij}\dot x^i\dot x^j-V=-H\,.
\eeqa

\subsection{Eisenhart Lift for Scalar Fields }

The Eisenhart lift of Lorentzian type can be extended to scalar field theories 
Eq. (\ref{action:scalar})  
in a similar manner as in the case of Riemannian type.

We introduce two additional vector fields 
$B^{\mu}$ and $C^{\mu}$ in place of new coordinates $u$ and $v$:
\beqa
I_L=\int d^4x\sqrt{-g}\left(\frac12 R-\frac12 g^{\mu\nu}k_{IJ}(\bphi)\p_{\mu}\phi^I\p_{\nu}\phi^J
+(\nabla_{\mu}B^{\mu})(\nabla_{\nu}C^{\nu})-V(\bphi)(\nabla_{\mu}B^{\mu})^2
\right)\,.
\label{action:scalar3}
\eeqa
The Einstein equation and the equations of motion of $\phi^{I},B^{\mu}$ and 
$C^{\mu}$ are
\beqa
&&R_{\mu\nu}-\frac12 g_{\mu\nu}R=k_{IJ}\p_{\mu}\phi^I\p_{\nu}\phi^J-\frac12 g_{\mu\nu}\left(g^{\alpha\beta}k_{IJ}\p_{\alpha}\phi^I\p_{\beta}\phi^J\right)\nonumber\\
&&~~~~~~~~~~~~~~~~~~+2B_{(\mu}\p_{\nu)}\pi_B-g_{\mu\nu}B^{\alpha}\p_{\alpha}\pi_B
+2C_{(\mu}\p_{\nu)}\pi_C-g_{\mu\nu}C^{\alpha}\p_{\alpha}\pi_C\nonumber\\
&&~~~~~~~~~~~~~~~~~~-g_{\mu\nu}\left(\pi_B\pi_C+V\pi_C^2\right)\,,
\label{einstein:scalar3}
\\
&&\Box\phi^I+\Gamma^{I}_{JK}g^{\mu\nu}\p_{\mu}\phi^J\p_{\nu}\phi^K-
\pi_C^2k^{IJ}\p_JV=0\,,
\label{eom:scalar3-1}
\\
&&\p_{\mu}\pi_C=0\,,
\label{eom:scalar3-2}\\
&&\p_{\mu}\pi_B=0\,,
\label{eom:scalar3-3}
\eeqa
where
\beqa
&&\pi_B=\nabla_{\mu}C^{\mu}-2V\nabla_{\mu}B^{\mu},\label{piB2}\\
&&\pi_C=\nabla_{\mu}B^{\mu}\label{piC}\,.
\eeqa
{}From Eq. (\ref{eom:scalar3-2}), we find $\pi_C$ is a constant. 
Setting $\pi_C=1$ in Eq. (\ref{eom:scalar3-1}) reproduces the equation of motion of the scalar fields Eq. (\ref{eom:scalar1}).  
Furthermore, from Eq. (\ref{eom:scalar3-3}), 
 $\pi_B$ is a constant. We find that setting $\pi_B=0$ in Eq. (\ref{einstein:scalar3}) reproduces the Einstein equation with the scalar fields 
Eq. (\ref{einstein:scalar1}).

\subsection{Scalar Field in the FLRW Universe: Lorentzian type}
\label{sec5}

We apply the formalism of the Lorentzian Eisenhart lift of scalar fields to 
the system of a single scalar field in the
FLRW universe (see also \cite{Cariglia:2018mos}). 

\subsubsection{Field Space}

For a flat FLRW universe $g_{\mu\nu}dx^{\mu}dx^{\nu}=-N(t)^2dt^2+a(t)^2d\x^2$, 
assuming that a scalar field $\phi$ and a vector $B^{\mu}$ are homogeneous, 
 the lifted system Eq. (\ref{action:scalar3}) reduces to the particle system 
 whose Lagrangian is
 \beqa
 {\cal L}=-\frac{3a}{N}\dot a^2+\frac{a^3}{2N}\dot\phi^2+\frac{1}{Na^3}\dot\chi_1\dot\chi_2-\frac{V}{Na^3}\dot\chi_1^2
 \equiv \frac12 G_{AB}\dot\varphi^A\dot\varphi^B\,,
 \eeqa
where $\chi_1\equiv Na^3B^0,  \chi_2\equiv Na^3C^0, \varphi^A=(a,\phi,\chi_1,\chi_2)$, and the field space metric $G_{AB}$ is given by
\beqa
G_{AB}=
\begin{pmatrix}
-\frac{6a}{N} & &  & \\
   & \frac{a^3}{N}& &  \\
  &  &-\frac{2V}{Na^3}&\frac{1}{Na^3} \\
 &  &  \frac{1}{Na^3}&  \\
\end{pmatrix}
\label{effective-metric2}
\,.
\eeqa
In terms of the conjugate momenta $p_a=-6a\dot a/N, 
p_{\phi}=a^3\dot\phi/N,  
p_{\chi_1}={\dot\chi_2}/{Na^3}-2{V\dot\chi_1}/{Na^3}, 
p_{\chi_2}={\dot\chi_1}/{Na^3}$,\footnote{$p_{\chi_1}$ and $p_{\chi_2}$ coincide with $\pi_B$ and $\pi_C$ in Eq. (\ref{piB2}) and Eq. (\ref{piC}), respectively.} the Hamiltonian becomes
\beqa
{\cal H}=\frac12G^{AB}p_Ap_B=N\left(-\frac{1}{12a}p_a^2+\frac{1}{2a^3}p_{\phi}^2+a^3p_{\chi_1}p_{\chi_2}+a^3Vp_{\chi_2}^2\right)\,,
\eeqa
from which the Hamiltonian constraint follows
\beqa
H= \frac{1}{2}\left(
-\frac{p_a^2}{6a}+\frac{p_\phi^2}{a^3}
+2a^3p_{\chi_1}p_{\chi_2}+2a^3Vp_{\chi_2}^2 \right)=0\,.
\label{H-constraint2}
\eeqa
Henceforth, we set $N=1$.  
The equations of motion of $\phi^A$ are derived from 
the canonical equations of motion,  $\dot\varphi^A=\frac{\p H}{\p p_A}, \dot p_A=-\frac{\p H}{\p \varphi^A}$ together with the Hamiltonian constraint Eq. \eqref{H-constraint2}.  Setting $p_{\chi_1}=0$ and $p_{\chi_2}=1$ reproduces 
the Einstein equations and the equation of motion of $\phi$ in a 
flat FLRW universe.

At first sight, $H$ seems to reproduce the Hamiltonian in 
\cite{Cariglia:2018mos} where  the Lorentzian lift of the scalar 
field system in the FLRW universe is performed. 
However, there is a big difference: $H$ contains the factor $a^{3}$ 
in front of $p_{\chi_1}p_{\chi_2}$.  Therefore, unlike the case of a particle,  $p_{\chi_1}$ cannot 
be regarded as  the Hamiltonian of the original scalar system. 

The Ricci and scalar curvatures are given by
\begin{align}
 R_{AB} =
\left(
\begin{array}{cccc}
-\frac{9}{2a^2} & &  &\\
 &-\frac34 & & \\
 & &\frac{-3V+2V''}{2a^6}&\frac{3}{4a^6}\\
 & & \frac{3}{4a^6}&0 \\
\end{array}
\right)
\end{align}
and
\begin{align}
R = \frac{9}{4a^3} \,.
\end{align}
Note that, unlike the case of the Riemannian type, the scalar curvature is 
independent of $V$ and never vanishes.
This leads to the fact that the field space of the Lorentzian type has less symmetry
than that of the Riemannian type.

In fact, since the metric does not depend on $\chi_1$ and $\chi_2$, we find 
two KVs $\eta_{(1)}=\p/\p\chi_1$ and $\eta_{(2)}= \p/\p\chi_2$. 
These are found to be the only CKVs satisfying 
the (C)KV equations irrespective of $V(\phi)$. 
$\xi_{(3)}$ is not a (C)KV on the field space of the Lorentzain type.\footnote{
For the potential $V(\phi)$ in Case II, we find that $\xi_{(3)}$ is hidden in KT. 
However, for  the potential $V(\phi)$  in the other cases, 
we could not find (C)KTs which involve $\xi_{(3)}$.}
Therefore, the formulation of the Eisenhart lift using two vector fields 
may not be useful to study the symmetry of the field space.





\bibliographystyle{apsrev4-1}
\bibliography{references}

\end{document}